\documentclass[a4paper,11pt]{article}
\pdfoutput=1
\usepackage{jheppub}

\usepackage{amssymb,amsmath}
\usepackage[normalem]{ulem}
\usepackage[utf8x]{inputenc}
\usepackage{slashed}
\usepackage{graphicx}
\usepackage{here}
\usepackage{color}
\usepackage{csquotes} 
\usepackage{comment} 
\usepackage{mathrsfs}
\usepackage{float}
\usepackage{ascmac}
\usepackage{multirow}
\usepackage{longtable}
\usepackage{bbm}

\usepackage{mathtools}

\usepackage{dcolumn}

\usepackage[italicdiff]{physics}

\usepackage[hang,small,bf]{caption}
\usepackage[subrefformat=parens]{subcaption}
\newcommand{\Ykr}[2]{Y^{(#1)}_{#2}}
\renewcommand{\vev}[1]{\left\langle{#1}\right\rangle}

\newcommand{\intZ}{\mathbb{Z}}
\newcommand{\CP}{\mathrm{CP}}
\newcommand{\hta}{\hat{\tau}}
\newcommand{\eps}{\epsilon}

\newcommand{\rta}{\mathrm{Re}\,\tau}
\newcommand{\ita}{\mathrm{Im}\,\tau}
\newcommand{\dta}{\delta \tau}

\makeatletter
\newcommand*\rel@kern[1]{\kern#1\dimexpr\macc@kerna}
\newcommand*\widebar[1]{%
  \begingroup
  \def\mathaccent##1##2{%
    \rel@kern{0.8}%
    \overline{\rel@kern{-0.8}\macc@nucleus\rel@kern{0.2}}%
    \rel@kern{-0.2}%
  }%
  \macc@depth\@ne
  \let\math@bgroup\@empty \let\math@egroup\macc@set@skewchar
  \mathsurround\z@ \frozen@everymath{\mathgroup\macc@group\relax}%
  \macc@set@skewchar\relax
  \let\mathaccentV\macc@nested@a
  \macc@nested@a\relax111{#1}%
  \endgroup
}
\makeatother

\numberwithin{equation}{section}

\preprint{
\begin{minipage}{5cm}
\small
\flushright
EPHOU-24-009\\
KYUSHU-HET-298 \\
CTPU-PTC-24-21
\end{minipage}}

\title{Moduli stabilization and light axion by Siegel modular forms}

\author{Shuta Funakoshi$^{1}$,} 
\author{Junichiro Kawamura$^{2}$,} 
\author{Tatsuo Kobayashi$^{3}$,} 
\author{Kaito Nasu$^{3}$, and} 
\author{Hajime Otsuka$^{1}$} 
\affiliation{
$^1$Department of Physics, Kyushu University, 744 Motooka, Nishi-ku, Fukuoka 819-0395, Japan}
\affiliation{
$^2$Center for Theoretical Physics of the Universe, Institute for Basic Science (IBS), Daejeon
34051, Korea}
\affiliation{
$^3$Department of Physics, Hokkaido University, Sapporo 060-0810, Japan}
\emailAdd{funakoshi.shuta@phys.kyushu-u.ac.jp}
\emailAdd{junkmura13@gmail.com}
\emailAdd{kobayashi@particle.sci.hokudai.ac.jp}
\emailAdd{k-nasu@particle.sci.hokudai.ac.jp}
\emailAdd{otsuka.hajime@phys.kyushu-u.ac.jp}

\abstract{
We discuss the stabilization of multiple moduli by utilizing Siegel modular forms 
in the framework of $Sp(2g,\mathbb{Z})$ modular invariant theories. 
We derive the stationary conditions at CP-conserving fixed points for a generic modular- and CP-invariant scalar potential. 
The stabilization of multiple moduli is explicitly demonstrated in $Sp(4,\mathbb{Z})$ and $Sp(6,\mathbb{Z})$ modular invariant scalar potentials. 
Furthermore, 
it turns out that there exists a light axion when the moduli are stabilized nearby a fixed point.
}

\makeatletter
\gdef\@fpheader{}
\makeatother

\begin{document}

\maketitle
\clearpage 

\section{Introduction}

Modular symmetry is of interest in several aspects of theoretical physics. 
Recently, the finite modular symmetries are regarded as a 
flavor symmetry of quarks and leptons \cite{Feruglio:2017spp}, where Yukawa couplings and masses are written by modular forms, and modular forms are 
multiplets of finite modular groups such as $S_3$, 
$A_4$, $S_4$, and $A_5$ \cite{Feruglio:2017spp,Kobayashi:2018vbk,Penedo:2018nmg,Novichkov:2018nkm}.
One can predict a 
peculiar observables in flavor physics, 
see e.g. Refs~\cite{Kobayashi:2023zzc,Ding:2023htn} for review.
Since the modular symmetry constrains the structure of four-dimensional (4D) effective 
action, it is also relevant to the cosmological history of the universe such as the inflation \cite{Kobayashi:2016mzg,Gunji:2022xig,Abe:2023ylh,Ding:2024neh,King:2024ssx,Casas:2024jbw}. 
It is important to discuss the potential of 
moduli fields determining the flavor structure of quarks and leptons 
as well as cosmological observables.

So far, there have been several attempts to stabilize the moduli fields 
in the framework of $SL(2,\mathbb{Z})$ modular symmetry \cite{Kobayashi:2019xvz,Kobayashi:2019uyt,Ishiguro:2020tmo,Ishiguro:2022pde,Novichkov:2022wvg,Kobayashi:2023spx,Knapp-Perez:2023nty,King:2023snq,Abe:2024tox,Leedom:2022zdm,Higaki:2024pql}~\footnote{ 
See for early works on moduli stabilization by modular invariant superpotential Refs.~\cite{Font:1990nt,Ferrara:1990ei,Cvetic:1991qm}.}.
In the top-down approach to realize the 4D modular symmetric models 
from string theory, 
the number of moduli fields is determined by the topological quantities of 
extra-dimensional space, i.e., the hodge numbers. 
A typical number is not one even for 
the complex structure or the K\"ahler structure of toroidal space. 
Indeed, there are ${\cal O}(100)$ moduli fields in Calabi-Yau compactifications. 
The 4D effective action enjoys the symplectic modular symmetry $Sp(2g,\mathbb{Z})$~\cite{Strominger:1990pd,Candelas:1990pi}, 
which includes the finite modular symmetry~\cite{Ishiguro:2020nuf,Ishiguro:2021ccl,Ishiguro:2024xph}~\footnote{
See also Refs.~\cite{Kikuchi:2023awe,Kikuchi:2023dow}.}.
In addition to the top-down approach, 
the symplectic modular symmetry was used in the bottom-up 
modular flavor models~\cite{Ding:2020zxw,Ding:2021iqp,RickyDevi:2024ijc,Ding:2024xhz}. 
Thus, 
it is quite important to study stabilization 
of those moduli fields associated with the $Sp(2g, \mathbb{Z})$ modular symmetry.

In this paper, 
we propose a method to stabilize the multiple moduli fields in the framework of 
$Sp(2g,\mathbb{Z})$ modular symmetric models with $g=2,3$. 
Compared with $SL(2,\mathbb{Z}) \simeq Sp(2,\mathbb{Z})$,
the moduli space in $Sp(2g, \mathbb{Z})$ 
has a rich structure 
and there are more CP-conserving fixed points in the moduli space. 
We show that the fixed points are stationary points in a generic modular- and CP-invariant scalar potential. 
We utilize the Siegel modular forms appearing in the modular invariant matter superpotential, 
so that the moduli fields are stabilized by the tree-level and/or 1-loop Coleman Weinberg 
potential~\cite{Kobayashi:2023spx,Higaki:2024jdk}. 
We explicitly demonstrate the moduli stabilization 
in $Sp(4,\mathbb{Z})$ and $Sp(6,\mathbb{Z})$ modular symmetric models. 
We find a case in which 
the modulus is stabilized near one of the fixed points 
and there remains a light mode compared with the other massive modes 
of the moduli fields. 
The light mode could be applicable for phenomenology, 
such as axion (like-particle) to explain the dark matter 
and/or resolve the strong CP problem.

This paper is organized as follows. 
In Sec. \ref{sec:2}, we show necessary conditions where 
a fixed point in the moduli space becomes stationary 
based on the modular and CP invariance of the modulus potential. 
For concreteness, 
we study the fixed points of $Sp(4,\mathbb{Z})$ modular symmetry in detail, 
although it is applicable to generic symplectic modular symmetries. 
In Sec.~\ref{sec:3}, 
we explicitly demonstrate the moduli stabilization under $Sp(4,\mathbb{Z})$ 
modular symmetry. 
Finally, Sec.~\ref{sec:con} is devoted to the conclusion. 
The singlet modular forms of $Sp(4,\mathbb{Z})$ 
are summarized in Appendix~\ref{app:sp4}, 
and then we discuss the moduli stabilization  
under $Sp(6,\mathbb{Z})$ modular symmetry in Appendix~\ref{app:sp6}.

\section{Stationary points in modular invariant scalar potentials}
\label{sec:2}
We review the $Sp(2g,\intZ)$ modular symmetry and CP symmetry in Sec.~\ref{sec:sp2gz}, 
and then we discuss the stationary points in the $Sp(2g,\intZ)$ 
and CP invariant scalar potential in Sec.~\ref{sec:extrema}.

\subsection{$Sp(2g,\intZ)$ modular symmetry and CP symmetry}
\label{sec:sp2gz}

We start with the Siegel modular group $\Gamma_g:=Sp(2g,\intZ)$ 
whose generators are given by
\begin{align}
    S=\left(
    \begin{array}{cc}
       0  &  \mathbbm1_g\\
      -\mathbbm1_g   & 0
    \end{array}\right)\,,
    \qquad
    T_m = \left(\begin{array}{cc}
      \mathbbm1_g & B_m \\
        0 & \mathbbm1_g
    \end{array}\right)\,.
\end{align}
Here, $\{B_m\}$ with $m=1,2,...,2g-1$ is a basis for the $g\times g$ integer symmetric matrices 
whose explicit form is shown later for $g=2,3$. 
The moduli fields transform as
\begin{align}
    \tau \xrightarrow{T_m} \tau +B_m\,,
    \qquad
    \tau \xrightarrow{S} -\tau^{-1}\,. 
    \label{eq:moduliTrf}
\end{align}
Here, $\tau$ is $g\times g$ matrix-valued, i.e. $\tau_{ij}=s_{ij}+it_{ij}$.
The moduli space of $\tau$ is described in the Siegel upper half plane~\cite{Siegel:1943}:
\begin{align}
    {\cal H}_g = \{ \tau \in GL(g,\mathbb{C}) 
\ | \ 
\tau^t = \tau,\quad {\rm Im}(\tau)>0\}, 
    \label{eq:Hg}
\end{align}
where $GL(g, \mathbb{C})$ is a set of $g\times g $ matrices with complex values. 
The modular transformation maps any value of $\tau$ to the following fundamental domain \cite{Siegel:1943}:
\begin{align}
    {\cal D}_g = 
    \left\{
    \tau \in {\cal H}_g \ \left|
    \quad 
    \begin{array}{l}
         h^T {\rm Im}(\tau) h \geq {\rm Im}(\tau)_{ii} \quad \forall h = (h_1,...,h_g)\in \mathbb{Z}^g 
          \quad {\rm for}\quad 
          1\leq i\leq g
         \\
         {\rm Im}(\tau)_{i,i+1}\geq 0\quad {\rm for}\quad 0\leq i \leq g-1
         \\
         |{\rm det}(C\tau+D)|\geq 1 
          \quad {\rm for}\quad 
          \forall \gamma
         =
         \begin{pmatrix}
             A & B \\
             C & D
         \end{pmatrix}\in \Gamma_g
         \\
         |{\rm Re}(\tau)_{ij}| \leq 1/2
    \end{array}
    \right.
    \right\}
    ,
\end{align}
where $\{h_1,...,h_g\}$ are coprime in the first line.
The Siegel modular group has finite modular groups. 
By introducing the principal congruence subgroup $\Gamma_g(n)$ of level $n$ with $n$ being a positive integer:
\begin{align}
\label{eq-defGgn}
    \Gamma_g(n) = \{ \gamma \in \Gamma_g ~|~ \gamma \equiv \mathbbm{1}_g\quad {\rm mod}~n\},
\end{align}
one can define the finite Siegel modular group as the quotient group $\Gamma_{g,n}:=\Gamma_g/\Gamma_g(n)$. 

The CP transformation of $\tau$:
\begin{align}
    \tau \xrightarrow{{\rm CP}} -\bar{\tau}    
\label{eq:CP1}
\end{align}
is regarded as an outer automorphism of the modular symmetry, leading to 
the extended modular symmetry $\Gamma_g \rtimes \mathbb{Z}_2^{\rm CP}$ \cite{Ishiguro:2020nuf,Ding:2021iqp}.\footnote{When $g=1$, the unification of CP and $SL(2,\mathbb{Z})$ modular symmetry was discussed in Refs. \cite{Baur:2019kwi,Novichkov:2019sqv}.} 
In the framework of modular invariant theories, the CP symmetry is conserved at the vacuum expectation value of $\tau$ 
up to a modular transformation of $\gamma \in \Gamma_g$:
\begin{align}
    -\bar{\tau}=\gamma \tau\,.
\label{eq:CPcon}
\end{align}
From the trivial transformation $\gamma = 1$, 
the CP symmetry is conserved on ${\rm Re}(\tau)=0$. 
Next, let us focus on $T_m$ transformation in $Sp(2g,\mathbb{Z})$. 
As explicitly shown later for $g=2,3$, one can choose the basis where $T_m$ transformation corresponds to the shift symmetry of the modulus $\tau_{ij}$, that is, $\tau_{ij}\rightarrow \tau_{ij}+1$. 
At the moment, 
we denote the modulus shifted by $B_m$ by $\tau_*$, i.e.  
\begin{align}
    \tau_{\ast} \xrightarrow{B_m} \tau_{\ast}+1.
\end{align}
By combining with $T_m$ and CP transformations for the point 
${\rm Re}(\tau_\ast)\equiv 0$ mod $1/2$: 
\begin{align}
    {\rm Re}(\tau_\ast) \xrightarrow{\rm{CP}} -{\rm Re}(\tau_\ast)  \xrightarrow{B_m} {\rm Re}(\tau_\ast)\,,
\end{align}
we find that ${\rm Re}(\tau_\ast)={1}/{2}$ are CP conserving lines. 
Hence, we conclude that CP is conserved at $|{\rm Re}(\tau)_{ij}| = 0, 1/2$. 
Other CP-conserving fixed points for $g=2,3$, i.e. $Sp(4,\mathbb{Z})$ and $Sp(6,\mathbb{Z})$ are discussed later.

\subsection{Extrema of the scalar potential}
\label{sec:extrema}

Let us consider the scalar potential which is invariant under the $\Gamma_g$ modular symmetry and the $\CP$ symmetry~\footnote{Stationary points in the $SL(2,\mathbb{Z})$ modular and CP-invariant potential were discussed in Ref.~\cite{Higaki:2024pql}.}. 
We can find the vanishing derivative of the moduli potential 
ensured by the $\Gamma_g \rtimes \mathbb{Z}_2^{\rm CP}$ symmetry 
as follows. 
We first show the simplest consequence from the $\CP$ symmetry alone, 
then eventually discuss those from the $\CP$ symmetry 
and $T$, $S$ and full modular invariance.  
The last discussion is the most general and includes the others.

\subsubsection*{$\CP$ invariance} 

From the $\CP$ invariance, the scalar potential obeys
\begin{align}
    V({\rm Re}(\tau), {\rm Im}(\tau)) = V(-{\rm Re}(\tau), {\rm Im}(\tau))\,.
\end{align}
This implies that $\rta = 0$ is extrema along the $\rta$ direction: 
\begin{align}
    \frac{\partial V}{\partial {\rm Re}(\tau_\ast)}\biggl|_{{\rm Re}(\tau_*)=0} =0\,,
\label{eq:Vcp}
\end{align}
where $\tau_\ast$ is any element of $\tau$.

\subsubsection*{$\CP$ and $T$ invariance}

Next, we consider $T_m$ transformation in $Sp(2g,\mathbb{Z})$. 
As discussed before, $T_m$ transformation corresponds to the shift symmetry of the modulus %
$\tau_{\ast} \rightarrow \tau_{\ast} + 1$. 
In what follows, we decompose $\tau_* := s + it$. 
Together with this modular transformation for $\tau_\ast$ and CP transformation for all the moduli \eqref{eq:CP1}, 
we obtain the following identity for the $T$- and $\CP$-invariant scalar potential:
\begin{align}
    V(s,t, {\rm Re}(\hta), {\rm Im}(\hta)) 
    = V(s+1, t, {\rm Re}(\hta), {\rm Im}(\hta)) 
    = V(-s-1, t, -{\rm Re}(\hta), {\rm Im}(\hta)), 
\end{align}
where $\hta$ is any element of $\tau$ other than $\tau_\ast$.
When we differentiate both sides with respect to $s={\rm Re}(\tau_\ast)$, 
we find 
\begin{align}
    \frac{\partial V}{\partial s}(s,t, {\rm Re}(\hta), {\rm Im}(\hta)) = -\frac{\partial V}{\partial s^\prime}(s^\prime, t,  
    -{\rm Re}(\hta), {\rm Im}(\hta)), 
\end{align}
with $s^\prime = -s-1$. 
This implies 
\begin{align}
    \frac{\partial V}{\partial s}\biggl|_{s=-1/2,{\rm Re}(\hta)=0} = 0, 
\end{align}
and is also vanishing at $s=1/2$ due to the $T_m$ symmetry. 
Applying this to all of $T_m$'s and combining with Eq.~\eqref{eq:Vcp},
we find 
\begin{align}
    \frac{\partial V}{\partial \rta_\ast} \biggl|_{\rta_\ast \equiv 0} = 0, 
\label{eq:Vs}
\end{align}
where $\equiv$ is understood as mod $1/2$. 
Thus, we conclude that ${\rm Re}(\tau)_{ij}\equiv 0$ (mod 1/2) 
for any $i,j$ is extrema of the modular $T$- and $\CP$-invariant scalar potential 
along the $\rta$ direction.

\subsubsection*{$\CP$, $S$ and $T$ invariance}

Next, we examine extrema ensured by  
the $S$ transformation in $Sp(2g,\mathbb{Z})$. 
For the $S$- and $\CP$-invariant scalar potential: 
\begin{align}
    V(\tau, \bar{\tau}) = V(-\bar{\tau}, -\tau)= V(\bar{\tau}^{-1}, \tau^{-1}),
\end{align}
we differentiate both sides with respect to $\tau$:
\begin{align}
\frac{\partial V}{\partial \tau_{ij}}(\tau, \bar{\tau}) &= 
+ \sum_{k,l} \frac{\partial u_{kl}}{\partial \tau_{ij}} \frac{\partial V}{\partial u_{kl}}(\bar{u}, u) 
=   - \sum_{k,l}(\tau^{-1})_{ki}(\tau^{-1})_{jl}\frac{\partial V}{\partial u_{kl}}(\bar{u}, u),
\end{align}
with $u=\tau^{-1}$ and $\bar{u}=(\bar{\tau})^{-1}$, 
where we use
\begin{align}
\frac{\partial}{\partial \tau_{ij}} ( \tau^{-1})_{kl} = - \sum_{m,n} (\tau^{-1})_{km}\frac{\partial \tau_{mn}}{\partial \tau_{ij}}(\tau^{-1})_{nl} = - (\tau^{-1})_{ki}(\tau^{-1})_{jl}.          
\end{align}
On the subspace where $\CP$ is conserved up to the $S$ transformation:
\begin{align}
   u  =  \tau^{-1} 
   = \bar{\tau} 
    \quad \leftrightarrow \quad 
        \bar{\tau} \tau = \tau \bar{\tau} = \mathbbm1_{g}, 
\label{eq:CPandS}
\end{align}
we find 
\begin{align}
    \frac{\partial V}{\partial \tau_{ij}} (\tau, \bar{\tau}) &=  - \sum_{k,l}(\tau^{-1})_{ki} (\tau^{-1})_{jl}   \frac{\partial V}{\partial u_{kl}} (\bar{u}, u)\biggl|_{u_{kl}=\bar{\tau}_{kl}, \bar{u}_{kl}=\tau_{kl}} 
    \nonumber\\
    &= - \sum_{k,l}\bar{\tau}_{ki}\,\bar{\tau}_{jl}   \frac{\partial V}{\partial u_{kl}} (\bar{u}, u)\biggl|_{u_{kl}=\bar{\tau}_{kl}, \bar{u}_{kl}=\tau_{kl}}.
    \label{eq:bartauij}
\end{align}
From Eq.~\eqref{eq:Vs},
\begin{align}
 \frac{\partial V}{\partial s_{ij}}(\tau, \bar{\tau})\Bigl|_{s_{ij}\equiv 0} 
    =  \left[
    \frac{\partial V}{\partial \tau_{ij}} (\tau, \bar{\tau})
    + \frac{\partial V}{\partial \bar{\tau}_{ij}}(\tau, \bar{\tau}) 
    \right]_{s_{ij}\equiv 0}= 0,  
    \label{eq:sij}
\end{align}
with $s_{ij} := \mathrm{Re}( \tau_{ij})$ 
and $\equiv$ is understood as mod $1/2$. 
By combining Eqs.~\eqref{eq:bartauij} and \eqref{eq:sij}, we obtain
\begin{align}
    \frac{\partial V}{\partial \tau_{ij}} 
=  \sum_{k,l}\bar{\tau}_{ik}\bar{\tau}_{jl}   \frac{\partial V}{\partial \tau_{kl}},
\end{align}
which can be rewritten as
\begin{align}
    \begin{pmatrix}
        \frac{\partial V}{\partial \tau_{11}}\\
        \frac{\partial V}{\partial \tau_{12}}\\
        \vdots\\ 
        \frac{\partial V}{\partial \tau_{nn}}
    \end{pmatrix}
    =
        \begin{pmatrix}
        \bar{\tau}_{11}\bar{\tau}_{11} & \bar{\tau}_{11}\bar{\tau}_{12} & \cdots & \bar{\tau}_{n1}\bar{\tau}_{1n}\\
        \bar{\tau}_{11}\bar{\tau}_{21} & \bar{\tau}_{11}\bar{\tau}_{22} & \cdots & \bar{\tau}_{n1}\bar{\tau}_{2n}\\
        \vdots & \vdots & \cdots & \vdots\\
        \bar{\tau}_{1n}\bar{\tau}_{n1} & \bar{\tau}_{1n}\bar{\tau}_{n2} & \cdots & \bar{\tau}_{nn}\bar{\tau}_{nn}\\
    \end{pmatrix}
    \begin{pmatrix}
        \frac{\partial V}{\partial \tau_{11}}\\
        \frac{\partial V}{\partial \tau_{12}}\\
        \vdots\\ 
        \frac{\partial V}{\partial \tau_{nn}}
    \end{pmatrix}
    .
    \label{eq:Vtauij_S}
\end{align}
Note that the above matrix is a $g(g+1)/2 \times g(g+1)/2$ matrix.
Hence, the scalar potential invariant under the modular and CP symmetry has the vanishing derivative 
where $\tau^{-1} = \overline{\tau}$
and  $s_{ij} \equiv 0$ (mod $1/2$) are satisfied  
if any eigenvalue of the above matrix is not unity. In particular, 
such point is a stationary point if all of the eigenvalues are not unity.

\subsubsection*{$\CP$ and modular invariance}

Finally, we extend the previous discussions to a generic modular transformation. 
We study the $\CP$-conserving points satisfying Eq.~\eqref{eq:CPcon} in the moduli space. 
For an arbitrary modular- and $\CP$-invariant scalar potential:
\begin{align}
    V(\tau, \bar{\tau}) = V(-\bar{\tau}, -\tau)= V(\gamma \tau, \overline{\gamma \tau}),
\end{align}
we differentiate both sides with respect to $\tau$:
\begin{align}
\frac{\partial V}{\partial \tau_{ij}}(\tau, \bar{\tau}) &=  \sum_{k,l} \frac{\partial u_{kl}}{\partial \tau_{ij}} \frac{\partial V}{\partial u_{kl}}(u, \bar{u}) =    \sum_{k,l}\frac{\partial (A\tau +B)_{km}(C\tau +D)^{-1}_{ml}}{\partial \tau_{ij}}\frac{\partial V}{\partial u_{kl}}(u, \bar{u})
\nonumber\\
&=    
\left( 
\sum_{k,m,l,p} 
A_{kp}\delta_{pi}\delta_{mj} (C\tau +D)^{-1}_{ml} +
\sum_{k,m,l} 
(A\tau +B)_{km} \frac{\partial (C\tau +D)^{-1}_{ml}}{\partial \tau_{ij}}\right)\frac{\partial V}{\partial u_{kl}}(u, \bar{u})
\nonumber\\
&=    \left( \sum_{k,l} A_{ki}(C\tau +D)^{-1}_{jl} - \sum_{k,m,l,n} (A\tau +B)_{km}(C\tau +D)^{-1}_{mn}C_{ni}(C\tau +D)^{-1}_{jl}\right)\frac{\partial V}{\partial u_{kl}}(u, \bar{u})
\nonumber\\
&=    \left( \sum_{k,l} A_{ki}(C\tau +D)^{-1}_{jl} 
- \sum_{k,l,n} u_{kn}C_{ni}(C\tau +D)^{-1}_{jl}\right)\frac{\partial V}{\partial u_{kl}}(u, \bar{u})
,
\end{align}
with $u=\gamma \tau = (A\tau +B)(C\tau +D)^{-1}$ and $\bar{u}=\overline{\gamma \tau}$, 
where we use
\begin{align}
\frac{\partial}{\partial \tau_{ij}} ( (C\tau +D)^{-1})_{kl} 
&= - \sum_{m,n} ((C\tau +D)^{-1})_{km}\frac{\partial (C\tau +D)_{mn}}{\partial \tau_{ij}}((C\tau +D)^{-1})_{nl} 
\nonumber\\
&= - \sum_{m,n,p} ((C\tau +D)^{-1})_{km}C_{mp} \delta_{pi}\delta_{nj}((C\tau +D)^{-1})_{nl}    
\nonumber\\
&= - \sum_{m} ((C\tau +D)^{-1})_{km}C_{mi}((C\tau +D)^{-1})_{jl} .    
\end{align}
When we insert $u= -\Bar{\tau}$ and $\bar{u} = -\tau$, we arrive at
\begin{align}
    \frac{\partial V}{\partial \tau_{ij}}(\tau, \bar{\tau}) 
    &= \left( \sum_{k,l} A_{ki}(C\tau +D)^{-1}_{jl} 
    - \sum_{k,l,n} u_{kn}C_{ni}(C\tau+D)^{-1}_{jl}\right)
    \frac{\partial V}{\partial u_{kl}}(u, \bar{u})
    \biggl|_{u_{kl}=-\bar{\tau}_{kl}, \bar{u}_{kl}=-\tau_{kl}}
    \nonumber\\
    &=  \left( \sum_{k,l} A_{ki}(C\tau +D)^{-1}_{jl} + \sum_{k,l,n} \bar{\tau}_{kn}C_{ni}(C\tau +D)^{-1}_{jl}\right)\frac{\partial V}{\partial (-\bar{\tau}_{kl})} (-\bar{\tau}, -\tau)
    \nonumber\\
    &=  -\left( \sum_{k,l} A_{ki}(C\tau +D)^{-1}_{jl} + \sum_{k,l,n} \bar{\tau}_{kn}C_{ni}(C\tau +D)^{-1}_{jl}\right)\frac{\partial V}{\partial \bar{\tau}_{kl}} (\tau, \bar{\tau})
    \nonumber\\
    &=: -\sum_{k,l} {\cal F}_{ij, kl}(\tau, \bar{\tau})\frac{\partial V}{\partial \bar{\tau}_{kl}} (\tau, \bar{\tau}),
    \label{eq:bartauij_general}
\end{align}
at $-\bar{\tau} = \gamma \tau$. Here, we use the identity for the CP-invariant scalar potential $V(\tau, \bar{\tau})= V(-\bar{\tau}, -
\tau)$:
\begin{align}
    \frac{\partial V}{\partial \bar{\tau}_{kl}} (\tau, \bar{\tau}) = -\frac{\partial V}{\partial (-\bar{\tau}_{kl})} (-\bar{\tau}, -\tau).
\end{align}
In the following, we discuss two cases: 
i) $s_{ij} \equiv 0\,({\rm mod}\,1/2)$ 
and 
ii) $s_{ij} \not\equiv 0\,({\rm mod}\,1/2)$.

\paragraph{i) $s_{ij} \equiv  0\quad ({\rm mod}\,1/2)$}\,\\

By combining Eqs.~\eqref{eq:bartauij_general} and \eqref{eq:sij}, we obtain 
\begin{align}
    \frac{\partial V}{\partial \tau_{ij}}(\tau, \bar{\tau}) =  \sum_{k,l} {\cal F}_{ij, kl}(\tau, \bar{\tau}) \frac{\partial V}{\partial \tau_{kl}} (\tau, \bar{\tau}),
    \label{eq:tauij}
\end{align}
which can be rewritten as
\begin{align}
    \begin{pmatrix}
        \frac{\partial V}{\partial \tau_{11}}\\
        \frac{\partial V}{\partial \tau_{12}}\\
        \vdots\\ 
        \frac{\partial V}{\partial \tau_{nn}}
    \end{pmatrix}
    =
        \begin{pmatrix}
        {\cal F}_{11,11} & {\cal F}_{11,12} & \cdots & {\cal F}_{11,nn}\\
        {\cal F}_{12,11} & {\cal F}_{12,12} & \cdots & {\cal F}_{12,nn}\\
        \vdots & \vdots & \cdots & \vdots\\
        {\cal F}_{nn,11} & {\cal F}_{nn,12} & \cdots & {\cal F}_{nn,nn}\\
    \end{pmatrix}
    \begin{pmatrix}
        \frac{\partial V}{\partial \tau_{11}}\\
        \frac{\partial V}{\partial \tau_{12}}\\
        \vdots\\ 
        \frac{\partial V}{\partial \tau_{nn}}
    \end{pmatrix}
    .
    \label{eq:Vtauij_general}
\end{align}
Again, the above matrix is $n$ by $n$ matrix with $n= g(g+1)/2$ being an independent number of moduli fields 
subject to $\tau^t = \tau$. 
Hence, when all the eigenvalues of the above matrix are not unity 
at a CP-conserving point satisfying Eq.~\eqref{eq:CPcon}, 
such points in the moduli space are stationary points of the modular- 
and $\CP$-invariant scalar potential.

\paragraph{ii) $s_{ij} \neq 0\quad ({\rm mod}\,1/2)$}\,\\

Together with \eqref{eq:bartauij_general} and its conjugate:
\begin{align}
    \frac{\partial V}{\partial \bar{\tau}_{ij}}(\tau, \bar{\tau}) 
    &= - \sum_{k,l} \overline{{\cal F}}_{ij, kl}(\tau, \bar{\tau})\frac{\partial V}{\partial \tau_{kl}} (\tau, \bar{\tau}),
\end{align}
we arrive at
\begin{align}
    \frac{\partial V}{\partial \tau_{ij}}(\tau, \bar{\tau}) 
    &=  \sum_{k,l,m,n} {\cal F}_{ij, kl}(\tau, \bar{\tau})\overline{{\cal F}}_{kl, mn}(\tau, \bar{\tau})\frac{\partial V}{\partial \tau_{mn}} (\tau, \bar{\tau}).
    \label{eq:bartauij_general_sij}
\end{align}
Thus, a generic $\CP$ conserving point is a stationary point 
if all of the eigenvalues of $\mathcal{F}\overline{\mathcal{F}}$ are not unity.

Here, we show a generic formula for the stationary conditions 
of moduli fields at CP-conserving points. 
For concreteness, we discuss the $\Gamma_2$ case in the next section 
and fixed points of the finite subgroup of $\Gamma_3$ in Appendix~\ref{app:sp6}.  
We note that the condition Eq.~\eqref{eq:tauij} is applicable 
to a general symplectic group $\Gamma_g$ with $g \ge 2$.

\section{Moduli stabilization under $Sp(4,\mathbb{Z})$}
\label{sec:3}
In Sec. \ref{sec:sp4}, 
we first briefly review the Siegel modular group $\Gamma_2=Sp(4,\mathbb{Z})$ 
and its finite subgroup, following Ref.~\cite{Ding:2020zxw}. 
We then demonstrate that all the fixed points 
of $\Gamma_2$ are stationary points using the results of the previous section.   
Next, we show a scenario which stabilizes the moduli fields 
by introducing a single matter in Sec.~\ref{sec:Sp4Zmodel1}. 
It will be turned out that there remain massless modes 
in this simplest setup. 
To stabilize the massless modes, 
we study a model with 
radiative potential induced by a vector-like matter in Sec. \ref{sec:Sp4ZCW}, 
and another singlet matter in Sec.~\ref{sec:Sp4Zmodel2}. 
Interestingly, there remains a light axion mode in the former case. 

\subsection{Brief review of $Sp(4,\mathbb{Z})$}
\label{sec:sp4}

\begin{table}[t]
    \centering
      \caption{\label{tab:Sp4Zfp}
      Fixed points of $\Gamma_2 = Sp(4,\mathbb{Z})$ and the generalized $\CP$ 
transformations 
       in the Siegel upper half plane ${\cal H}_2$ \cite{Ding:2020zxw,Ding:2021iqp}. 
     Here, we define $\zeta := e^{2\pi i/5}$, $\eta:= (1+i 2\sqrt{2})/3$ 
     and $\omega := e^{2\pi i/3}$.}
    \begin{tabular}{|c|c|c|}\hline
      Fixed points   &  Generators of the stabilizer group
      & 
      $\CP$ transformations\\
      \hline
$\left(
\begin{array}{cc}
  \zeta & \zeta + \zeta^{-2}  \\
  \zeta + \zeta^{-2}   & -\zeta^{-1}
\end{array}
\right)$
&
$\left(
\begin{smallmatrix}
  0 & -1 & -1 & -1  \\
  0 & 0 & -1 & 0  \\
  0 & 0 & 0 & -1  \\
  1 & 0 & 0 & 1  \\    
\end{smallmatrix}
\right)$
&
$((ST_3)^3T_3)^{-1} \mathcal{CP}$
\\
\hline
$\left(
\begin{array}{cc}
  \eta & \frac{1}{2}(\eta -1)  \\
  \frac{1}{2}(\eta -1) & \eta
\end{array}
\right)$
& 
$\left(
\begin{smallmatrix}
  0 & 1 & 0 & 0  \\
  1 & 0 & 0 & 0  \\
  0 & 0 & 0 & 1  \\
  0 & 0 & 1 & 0  \\
\end{smallmatrix}
\right)$
,
$\left(
\begin{smallmatrix}
  -1 & 1 & 1 & 0  \\
  1 & 0 & 0 & 1  \\
  -1 & 0 & 0 & 0  \\
  1 & -1 & 0 & 1 \\
\end{smallmatrix}
\right)$
&
$ST_3(T_1T_2)^{-1}S\,{\cal CP}$
\\
\hline
$\left(
\begin{array}{cc}
  i & 0  \\
  0 & i
\end{array}
\right)$
& 
$\left(
\begin{smallmatrix}
  0 & 0 & 1 & 0  \\
  0 & -1 & 0 & 0  \\
  -1 & 0 & 0 & 0  \\
  0 & 0 & 0 & -1  \\
\end{smallmatrix}
\right)$
,
$\left(
\begin{smallmatrix}
  0 & 0 & -1 & 0  \\
  0 & 0 & 0 & 1  \\
  1 & 0 & 0 & 0  \\
  0 & -1 & 0 & 0 \\
\end{smallmatrix}
\right)$
,
$\left(
\begin{smallmatrix}
  0 & 1 & 0 & 0  \\
  1 & 0 & 0 & 0  \\
  0 & 0 & 0 & 1  \\
  0 & 0 & 1 & 0 \\
\end{smallmatrix}
\right)$
&
${\cal CP}$
\\
\hline
$\left(
\begin{array}{cc}
  \omega & 0  \\
  0 & \omega
\end{array}
\right)$
& 
$\left(
\begin{smallmatrix}
  0 & 0 & 0 & -1  \\
  1 & 0 & 1 & 0  \\
  0 & 1 & 0 & 1  \\
  -1 & 0 & 0 & 0  \\
\end{smallmatrix}
\right)$
,
$\left(
\begin{smallmatrix}
  0 & 1 & 0 & 0  \\
  1 & 0 & 0 & 0  \\
  0 & 0 & 0 & 1  \\
  0 & 0 & 1 & 0 \\
\end{smallmatrix}
\right)$
,
$\left(
\begin{smallmatrix}
  0 & 0 & 1 & 0  \\
  0 & 0 & 0 & -1  \\
  -1 & 0 & -1 & 0  \\
  0 & 1 & 0 & 1 \\
\end{smallmatrix}
\right)$
&
$(T_1T_2)^{-1}  {\cal CP}$
\\
\hline
$\frac{i}{\sqrt{3}}
\left(
\begin{array}{cc}
  2 & 1  \\
  1 & 2
\end{array}
\right)$
& 
$\left(
\begin{smallmatrix}
  0 & 0 & 0 & 1  \\
  0 & 0 & 1 & 1  \\
  1 & -1 & 0 & 0  \\
  -1 & 0 & 0 & 0  \\
\end{smallmatrix}
\right)$
,
$\left(
\begin{smallmatrix}
  0 & 0 & 1 & 1  \\
  0 & 0 & 1 & 0  \\
  0 & -1 & 0 & 0  \\
  -1 & 1 & 0 & 0 \\
\end{smallmatrix}
\right)$
,
$\left(
\begin{smallmatrix}
  0 & 0 & 0 & 1  \\
  0 & 0 & -1 & 0  \\
  0 & -1 & 0 & 0  \\
  1 & 0 & 0 & 0 \\
\end{smallmatrix}
\right)$
&
${\cal CP}$

\\
\hline
$\left(
\begin{array}{cc}
  \omega & 0  \\
  0 & i
\end{array}
\right)$
& 
$\left(
\begin{smallmatrix}
  0 & 0 & 1 & 0  \\
  0 & 0 & 0 & 1  \\
  -1 & 0 & -1 & 0  \\
  0 & -1 & 0 & 0  \\
\end{smallmatrix}
\right)$
&
$T_1^{-1}{\cal CP}$

\\
\hline
    \end{tabular}
\end{table}

Let us start with the Siegel modular group $\Gamma_2=Sp(4,\mathbb{Z})$, 
which can be generated by four elements:
\begin{align}
    T_1 = \left(\begin{array}{cc}
      \mathbbm1_2 & B_1 \\
        0 & \mathbbm1_2
    \end{array}\right)\,,
    \quad
    T_2=\left(
    \begin{array}{cc}
      \mathbbm{1}_2   &  B_2\\
        0 & \mathbbm{1}_2
    \end{array}\right)\,,
    \quad
    T_3=\left(\begin{array}{cc}
       \mathbbm1_2  & B_3 \\
         0& \mathbbm1_2
    \end{array}\right)\,,
    \quad
    S=\left(
    \begin{array}{cc}
       0  &  \mathbbm1_2\\
      -\mathbbm1_2   & 0
    \end{array}\right)
    ,
\end{align}
with
\begin{align}
    B_1=\left(\begin{array}{cc}
      1   &  0\\
      0   & 0
    \end{array}\right)\,,
    \quad
    B_2=\left(\begin{array}{cc}
      0   & 0 \\
      0   & 1
    \end{array}\right)\,,
    \quad
    B_3=\left(\begin{array}{cc}
      0   &  1\\
      1   & 0
    \end{array}\right).
\end{align}
The moduli fields transform as in Eq.~\eqref{eq:moduliTrf}.

Let us examine the fixed point of $\Gamma_2 = Sp(4,\mathbb{Z})$ in the Siegel upper half plane. 
As shown in Table~\ref{tab:Sp4Zfp}, 
there are six fixed points classified by Refs.~\cite{gottschling1961fixpunkte,gottschling1961fixpunktuntergruppen,gottschling1967uniformisierbarkeit} 
which are defined as
\begin{align}
    h_\ast\,\tau = \tau.
\end{align}
The elements $h_\ast$ constitute a subgroup $H$ of $\Gamma_2$. 
Since $\tau$ is invariant under $-\mathbbm{1}_{4}$, one can define the quotient group $\bar{H}=H/\{\pm \mathbbm{1}_{4}\}$, defined as a stabilizer group. 
We list this stabilizer group $\bar{H}$ for the six fixed points in Table \ref{tab:Sp4Zfp}. 
Note that $\tau$ is given by a $2\times 2$ matrix in the Siegel upper half plane ${\cal H}_2$, 
where the number of independent moduli is three. 
We find that when the three fixed points of $\Gamma_2$ satisfy Eq.~\eqref{eq:CPandS}, 
i.e. diagonal $\tau$ in the left column of Table~\ref{tab:Sp4Zfp}, 
all the eigenvalues of the matrix in Eq.~\eqref{eq:Vtauij_S} are not unity.
For instance, the point $\tau = \mathrm{diag}(i,i)$ has eigenvalue $(0,0,2)$. 
Furthermore, all the eigenvalues of the matrices 
in Eqs.~\eqref{eq:Vtauij_general} and \eqref{eq:bartauij_general_sij} 
are not unity at the other non-diagonal fixed points of $\tau$ with 
$s_{ij}\equiv0\,({\rm mod}\,1/2)$ and $s_{ij}\neq 0$ in Table \ref{tab:Sp4Zfp}, respectively. 
For instance, for the fixed point in the first column of Table \ref{tab:Sp4Zfp}, 
the matrix in Eq. \eqref{eq:Vtauij_general} is represented as
\small
\begin{align}
        \begin{pmatrix}
        {\cal F}_{11,11} & {\cal F}_{11,12} & {\cal F}_{11,22}\\
        {\cal F}_{12,11} & {\cal F}_{12,12} & {\cal F}_{12,22}\\
        {\cal F}_{22,11} & {\cal F}_{22,12} & {\cal F}_{22,22}
    \end{pmatrix}
    =
    \begin{pmatrix}
-0.854102 + 0.726543 i & 0.726543 i & 0.381966 \\
- 0.726543 i & 0.690983 - 0.951057 i & -0.5 - 0.363271 i \\
-0.381966 & -0.5 - 0.363271 i & -1
    \end{pmatrix}
    ,
\end{align}
\normalsize
whose eigenvalues are not unity. 
Hence, this fixed point is a stationary point. 
By analyzing the other cases, we verify that all the fixed points of $\Gamma_2$ are stationary points in a generic modular- and CP-invariant scalar potential.

The Siegel modular group allows the existence of finite modular groups. 
In this paper, we focus on the finite Siegel modular group 
$\Gamma_{2,2} := \Gamma_2/\Gamma_2(2)$, 
where $\Gamma_g(n)$ is defined in~Eq.~\eqref{eq-defGgn}. 
$\Gamma_{2,2}$ is isomorphic to $S_6$ 
which are generated by $F_1= ST_3$ and $F_2=T_1$ satisfying 
\begin{align}
    (F_2)^2= (F_1)^6= (F_2F_1)^5= (F_2 F_1^3)^4 = (F_2F_1^4F_2F_1^2)^2=1.
\end{align}
Note that the $S_6$ symmetry can be realized for a generic point of the moduli space of $\tau$:
\begin{align}
    \tau=
\begin{pmatrix}
    \tau_{11} & \tau_{12} \\ \tau_{21} & \tau_{22}
\end{pmatrix}
\equiv
    \left(\begin{array}{cc}
      \tau_1   &  \tau_3\\
      \tau_3   & \tau_2
    \end{array}\right)\, .
\end{align}
When we further restrict to $\tau_1=\tau_2$, 
the $S_6$ symmetry reduces to $S_4\times \mathbb{Z}_2$ 
known as the stabilizer group. 
The finite modular group $S_4\times \intZ_2$ can be generated by 
\begin{align}
    \mathcal{S}=T_1 T_2,
\quad 
\mathcal{T}=(G_3 G_2)^4 = (ST_3)^4,
\quad 
\mathcal{V}=(S T_3)^3,
\end{align}
satisfying the following relations:
\begin{align}
    \mathcal{S}^2=\mathcal{T}^3=(\mathcal{S}\mathcal{T})^4=1,
\quad 
{\cal V}^2 = 1, 
\quad 
\mathcal{S}\mathcal{V}=\mathcal{V}\mathcal{S},
\quad
\mathcal{T}\mathcal{V}=\mathcal{V}\mathcal{T}.
\end{align}
Here the generators of $S_4$ and $\mathbb{Z}_2$ 
are given by $\{{\cal S}$, ${\cal T}\}$ and ${\cal V}$, 
respectively. 
In the following, 
we demonstrate the stabilization of $\tau_1$ and $\tau_3$ 
in the finite modular symmetric model $S_4 \times \intZ_2$
with fixing $\tau_1 = \tau_2$. 
This would be achieved by a certain stabilization 
of the full modular space of $\Gamma_{2,2}$,
but it is a subject of our future work.

\subsection{Moduli stabilization by single modular form}
\label{sec:Sp4Zmodel1}

In this section, we study the supersymmetric effective action in the framework of 
$S_4\times \intZ_2$ modular symmetry. 
In addition to the moduli fields, we introduce a chiral superfield $X$ whose effective K\"ahler potential and superpotential in units of the reduced Planck mass are described by
\begin{align}
    K &= -h \ln \left( \det (-i\tau + i \Bar{\tau})\right) + \frac{|X|^2}{\det (-i\tau + i \Bar{\tau})^{k_X}}\,,
    \nonumber\\
    W &= \Lambda_X^2 Y_{\mathbf{1}}^{(k)}(\tau) X,
    \label{eq-Wsingle}
\end{align}
where $h$ is a positive integer, $\Lambda_X$ is a mass scale, and $Y_{\mathbf{1}}^{(k)}(\tau)$ is a singlet modular form 
with weight $k = k_X - h$ under $S_4 \times \mathbb{Z}_2$
so that $e^K \abs{W}^2$ is modular invariant. 
The explicit forms of the singlet modular forms are shown 
in Appendix \ref{app:sp4}. 
The scalar potential is given by
\begin{align}
    V = e^K \left( K^{I\bar{J}}D_I W D_{\bar J}\bar{W} - 3 |W|^2 \right),
\end{align}
where $D_I W := \partial_I W + W\partial_I K$ 
with $I=\{\tau_i, X\}$ and $K^{I\bar{J}}$ 
is the inverse of the K\"ahler metric.

In general, supersymmetric point satisfying $D_I W = 0$ is a minimum 
of the scalar potential. In our case, it is given by  
\begin{align}
    \partial_X W &= \Lambda_X^2 Y_{\mathbf{1}}^{(k)} = 0,
   \quad 
    \partial_{\tau_i} W = \Lambda_X^2 (\partial_{\tau_i}Y_{\mathbf{1}}^{(k)}) X = 0,
    \quad 
    W  = 0. 
\label{eq:SUSY}
\end{align}
At this point, 
the matter $X$ and moduli $\tau_{1,3}$ may be stabilized at 
\begin{align}
    \langle Y_{\mathbf{1}}^{(k)} (\tau)\rangle = \langle X \rangle = 0.
\label{eq:vev_sp4z1}
\end{align}
Hence, zero points of the modular form $Y_{\mathbf{1}}^{(k)}$ are minima.
Since the modular forms tend to have zero points at the fixed points, 
those are candidates of the minimum of the potential.

In general, 
the modular forms are zero at 
a fixed point $\tau_{h^*}$ of the $h_*$ transformation in the stabilizer group 
if the automorphic factor 
$\det(C_h \tau_h + D_h)$ is not unity 
since 
\begin{align}
   Y_{\mathbf{1}}^{(k)}(\tau_{h^*}) = 
   Y_{\mathbf{1}}^{(k)}(h_\ast \tau_{h^*}) 
   = \bigl[{\rm det}(C_{h^*} \tau_{h^*} +D_{h^*})]^{k} 
  Y_{\mathbf{1}}^{(k)}(\tau_{h^*}),
\label{eq:Ymodulartrf}    
\end{align}
where $C_{h^*}$ and $D_{h^*}$ correspond to 
$C$ and $D$ of the $h^*$ transformation, respectively. 
For $k=2$, we find that only at the fixed point:
\begin{align}
    \langle \tau_1\rangle = \omega,\qquad \langle \tau_3\rangle = 0
\label{eq:vev_sp4z_24a1}
\end{align}
the modular form is zero among the fixed points in Table~\ref{tab:Sp4Zfp}. 
Note that the moduli space enjoys a residual $\mathbb{Z}_2$ symmetry around $\tau_3=0$. 
It can be understood by the following configuration of moduli fields:
    \begin{align}
        \tau= 
        \begin{pmatrix}
            \tau_1 & 0\\
            0 & \tau_1
        \end{pmatrix}
        ,
    \end{align}
which is invariant under the $\mathbb{Z}_2$ subgroup of $\Gamma_2$ 
\cite{Ding:2020zxw}:
    \begin{align}
        h_\ast\,\tau = \tau,
    \end{align}
    with
    \begin{align}
        h_\ast =
        \begin{pmatrix}
            \sigma_3 & 0\\
            0 & \sigma_3\\ 
        \end{pmatrix}
    .
    \end{align}
Here, $\sigma_3$ is a Pauli matrix. 
When we turn on the modulus $\tau_3$, the above $\mathbb{Z}_2$ transformation flips the sign of $\tau_3$:
    \begin{align}
    \label{eq-residualZ2}
        \tau_3 \rightarrow - \tau_3,
    \end{align}
at $\tau_3=0$. 
Hence, there exists the residual $\mathbb{Z}_2$ symmetry around $\tau_3=0$, and Siegel modular forms respect this symmetry.

It turns out that the modular forms up to the weight six listed in Appendix \ref{app:sp4} except for $Y_{\mathbf{1}_c}^{(6)}$ are zero at the fixed point \eqref{eq:vev_sp4z_24a1}. For more details about zero points of $Y_{\mathbf{1}}^{(k)}$, see Appendix \ref{app:sp4}. 
Hence, our scenario is also applicable to these modular forms. 
We note that the modular form can be zero 
at generic points other than the fixed points, 
as we confirmed numerically. 
Physics around such a generic zero point would be interesting, 
but we focus on those at the fixed points 
so that we can discuss consequences of the residual symmetries.

The potential generated by the superpotential Eq.~\eqref{eq-Wsingle} 
can give masses for one direction of the moduli  superfields 
$\tau_1$ and $\tau_3$, so there remains a massless direction. 
From the expansion of the superpotential at the fixed point up to the second order of $\delta X= X- \langle X\rangle$ and $\delta \tau_{1,3} = \tau_{1,3} - \langle\tau_{1,3}\rangle$:
\begin{align}
    W = \Lambda_X^2 \delta X \left( \frac{\partial Y_{\mathbf{1}}^{(2)}}{\partial \tau_1} \delta \tau_1 +
    \frac{\partial Y_{\mathbf{1}}^{(2)}}{\partial \tau_3} \delta \tau_3 \right).
\end{align}
Since the supersymmetric mass matrix for 
$(\delta X, \delta \tau_1, \delta \tau_3)$ is rank-2, 
there is a massless direction in the global supersymmetry limit.
We can also see this in the scalar potential,  
\begin{align}
    V_F \sim \abs{\frac{\partial W}{\partial \delta X}}^2 \propto \abs{\frac{\partial Y_{\mathbf{1}}^{(2)}}{\partial \tau_1} \delta \tau_1 +
    \frac{\partial Y_{\mathbf{1}}^{(2)}}{\partial \tau_3} \delta \tau_3}^2.   
\end{align}
So, in general, 
only a linear combination of $\tau_1$ and $\tau_3$ can be stabilized. 
Specifically, 
$\partial_{\tau_3} Y_{\mathbf{1}}^{(2)}$ vanishes at the fixed point \eqref{eq:vev_sp4z_24a1} 
as analytically shown in Appendix~\ref{app:sp4}, 
and thus 
$\tau_3$ corresponds to the massless mode. 
We can see this feature by expanding the potential 
around the fixed point $(\tau_1, \tau_3) = (\omega,0)$, 
\begin{align}
\label{eq-VXseries}
V =&\ \Lambda_X^4 \left\{ 
(2\ita_1)^2 - (2\ita_3)^2 
\right\}^2 \abs{\Ykr{2}{1}(\tau)}^2 
\simeq  
\lambda_X 
\Lambda_X^4  
\abs{\delta \tau_3}^4
\Biggl( 
1 - f_X(\phi) \abs{\delta \tau_3}^2 
 \Biggr)
, 
\end{align}
where 
\begin{align}
\lambda_X :=&\ \frac{9}{4}
\abs{\vev{\partial_{\tau_3}^2 \Ykr{2}{1}}}^2,  
\\
f_X(\phi) :=&\ \frac{1}{6} \left\{ 
16 (1-c_\phi^2) + 
\vev{\frac{\partial_3^4 \Ykr{2}{1}}{\partial_3^2 \Ykr{2}{1}}}
(1-2c_\phi^2) 
\right\},
\end{align}
where $\order{(\delta \tau_3)^8}$ is neglected and 
$\delta \tau_3 =: \abs{\delta \tau_3} e^{i\phi}$, $\cos{\phi}=:c_{\phi}$. 
Here, $\vev{\cdots}$ is evaluated at the fixed point. 
Note that the odd derivatives of the modular forms 
at $\tau_3 = 0$ are vanishing. 
This feature can also be understood by the residual $\mathbb{Z}_2$ 
symmetry at the fixed point. 
Thus, $\delta \tau_3$ is massless at the fixed point 
since the potential is proportional to $\abs{\delta\tau_3}^4$.

In order to stabilize the massless mode, we consider the two scenarios: 
\begin{enumerate}
    \item{ 
    We add a vector-like mass term which generates a radiative potential. 
    }

    \item{We add the second matter coupling which generates a tree-level potential. } \end{enumerate}
The first scenario is discussed in Sec.~\ref{sec:Sp4ZCW}, 
and then the second one is discussed in Sec.~\ref{sec:Sp4Zmodel2}.

\subsection{Radiative stabilization of $\tau_3$ and light axion}
\label{sec:Sp4ZCW}

We introduce a vector-like pair $(Q, \bar{Q})$ 
under the Standard Model (SM) gauge group to stabilize the modulus $\tau_3$. 
The K\"ahler potential and superpotential are given by
\begin{align}
K_Q &= 
    \frac{Q^\dagger Q}{{\rm det}(-i\tau + i \Bar{\tau})^{k_{Q}}} 
    +\frac{\overline{Q}^\dagger \overline{Q}}
          {{\rm det}(-i\tau + i \Bar{\tau})^{k_{\overline{Q}}}}\,,
    \nonumber\\
    W_Q &= 
    \Lambda_{Q} \Ykr{k_{Y_Q}}{r_Q}(\tau) \overline{Q} Q, 
\end{align}
where $\Lambda_{Q}$ is a mass scale.
The modular weights satisfy 
$k_{Q} + k_{\overline{Q}} = k_{Y_Q} +h$, and the representation $r_Q$ is chosen such that 
the superpotential $W_Q$ is modular invariant up to the automorphic factor. 
The vector-like mass induces 
the one-loop Coleman-Weinberg (CW) potential: 
\begin{align}
    V_{\rm CW} =&\ \frac{1}{32\pi^2} \biggl[ (m_0^2 + m_Q^2(\tau))^2 \left\{ \ln \left(\frac{m_0^2 + m_Q^2(\tau)}{\mu^2}\right) -\frac{3}{2} \right\}
    \notag \\ 
    & \hspace{3.5cm}
    - (m_Q^2(\tau))^2 \left\{ \ln \left(\frac{m_Q^2(\tau)}{\mu^2}\right) - \frac{3}{2} \right\} \biggl],
\end{align}
where $\mu$ is the renormalization scale in the $\overline{{\rm MS}}$ scheme,
$m_0^2$ is the soft supersymmetry breaking scalar mass squared. 
In this paper, we assume that $m_0$ is independent of the modulus $\tau$
for simplicity~\footnote{
The soft scalar masses may depend on a modulus 
through a modular form in the moduli-mediated supersymmetry-breaking scenario, as pointed out in Ref.~\cite{Kikuchi:2022pkd}.}.
Here, $m_Q^2(\tau)$ is the moduli-dependent supersymmetric mass:
\begin{align}
    m_Q^2(\tau) := \Lambda_Q^2 \; 
{\rm det}(-i\tau + i \Bar{\tau})^{k_{Q}} \abs{\Ykr{k_{Y_Q}}{r_{Q}}(\tau)}^2.
    \label{eq:mQ2}
\end{align}

We analyze the whole scalar potential: 
\begin{align}
    V_{\rm tot} = V_{\rm tree} + V_{\rm CW},
\label{eq:Vtot}
\end{align}
where the tree-level potential is given by 
    \begin{align}
    V_{\rm tree} =&\ \Lambda_X^4  \det(-i\tau+i\tau^\dag)^2 \abs{\Ykr{2}{1}(\tau)}^2.
\end{align}
We assume that the tree-level potential dominates over the CW potential, 
so that the moduli are stabilized near the fixed point 
$\tau_0 := (\tau_1, \tau_3) = (\omega,0)$.
In particular, $\tau_1$ is stabilized at the fixed point, $\vev{\tau_1} = \omega$ due to the positive mass square from the tree-level potential. 
Up to $k\le 6$, there is only one singlet modular form 
which is non-zero at the fixed point, 
namely $\Ykr{6}{1_c}$ defined in Appendix~\ref{app:sp4}.  
Hence we study the case with $k_{Y_Q}=6$ and $r_Q = 1_c$.
Throughout this paper, we assume $m_0^2 \ll m_Q^2(\tau_0)$, 
so the CW potential is approximately given by 
\begin{align}
    V_{\mathrm{CW}} =&\  
    \frac{m_0^2 m_Q^2(\tau)}{16\pi^2} 
    \log \left(\frac{m_Q^2(\tau)}{e\mu^2}\right) + \order{m_0^4} 
    \\ \notag 
    = &\ \frac{m_0^2 m_Q^2(\tau_0)}{16\pi^2} 
\Biggl[
(-1+L_Q) +\left\{
-8 + 8 c_\phi^2 + 
\vev{\frac{\partial_3^2 \Ykr{6}{1_c}}{\Ykr{6}{1_c}}} 
 (2c^2_{\phi}-1) 
\right\} 
L_Q \abs{\delta \tau_3}^2 
\\ \notag 
&\ + 
 \Biggl\{
\frac{16}{3} (1-c_\phi^2)^2 (6+5L_Q) 
+ 
8(1+L_Q)\left(1-3c_\phi^2 + 2c_\phi^4\right)  
\vev{\frac{\partial_3^2 \Ykr{6}{1_c}}{\Ykr{6}{1_c}}} 
\\ \notag 
&\ + \frac{1}{4}
\left(2-8c_\phi^2 + 8 c_\phi^4 + L_Q  \right) 
\vev{\frac{\partial_3^2 \Ykr{6}{1_c}}{\Ykr{6}{1_c}}}^2 
+ \frac{L_Q }{12}
\left(1-8c_\phi^2 + 8 c_\phi^4  \right)  
\vev{\frac{\partial_3^4 \Ykr{6}{1_c}}{\Ykr{6}{1_c}}} 
\Biggr\}
\abs{\delta \tau_3}^4
\Biggr] 
\\ \notag 
&\ + \order{(\delta \tau_3)^6, m_0^4}, 
\end{align}
where  $L_Q := \log ({m_Q^2(\tau_0)}/{\mu^2})$,  
$\phi := \mathrm{Arg}(\delta \tau_3)$ and $c_\phi := \cos\phi$.    
We find 
\begin{align}
\vev{\frac{\partial_3^2 \Ykr{6}{1_c}}{\Ykr{6}{1_c}}} = -4,   
\quad 
\vev{\frac{\partial_3^4 \Ykr{6}{1_c}}{\Ykr{6}{1_c}}} 
= 193.6, 
\quad 
\vev{\frac{\partial_3^4 \Ykr{2}{1}}{\Ykr{2}{1}}} 
= -37.79, 
\end{align}
at the fixed point. 
Note that the first ratio is an integer value, and hence, remarkably, 
the $\order{\abs{\delta \tau_3}^2}$ term is independent of the phase $\phi$.  
This fact will make the hierarchy between the masses of the radial and phase modes.

For simplicity, we choose the renormalization scale $\mu$ at $L_Q = 1$, and 
then the full potential is given by 
\begin{align}
 V =   V_{\mathrm{tree}} + V_{\mathrm{CW}}
    \simeq &\ \lambda_X \Lambda_X^4 
    \Biggl[
\abs{\delta \tau_3}^4
\left(1-f_X(\phi) \abs{\delta \tau_3}^2\right) 
-2\epsilon_Q \abs{\delta \tau_3}^2
\left(
1 - f_Q(\phi) \abs{\delta \tau_3}^2 
\right)
    \Biggr], 
\end{align}
where 
\begin{align}
    \eps_Q := \frac{m_0^2 m_Q^2(\tau_0)}
                   {8\pi^2 \lambda_X \Lambda_X^4}, 
                \quad 
   f_Q(\phi) := \frac{1}{3}(5+32c_\phi^2 -28 c_\phi^4) 
     + 
\frac{1}{48}\left(1-8c_\phi^2 + 8c_\phi^4 \right)
\vev{\frac{\partial_3^4 \Ykr{6}{1_c}}{\Ykr{6}{1_c}}} . 
\end{align}
Assuming $\eps_Q \ll 1$, the radial direction is stabilized at 
\begin{align}
    \abs{\dta_3}^2 = \eps_Q + \order{\eps_Q^2}, 
\end{align}
thus the deviation from the fixed point is $\order{\eps_Q^{1/2}}$.

\begin{figure}[t]
\begin{minipage}{0.49\hsize}
  \begin{center}
  \includegraphics[height=60mm]{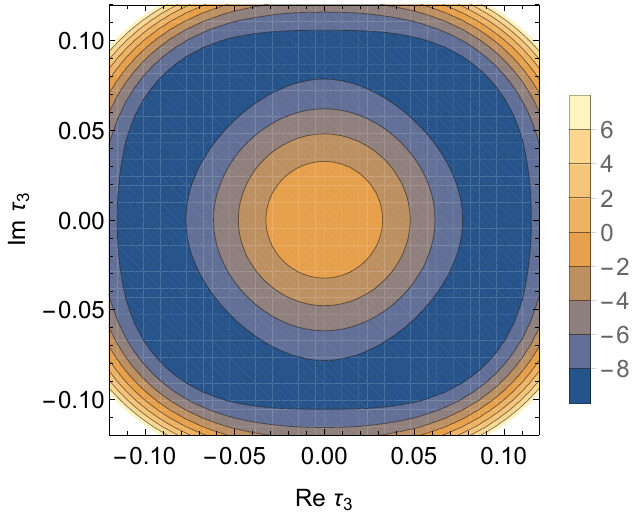}
  \end{center}
 \end{minipage}
 \begin{minipage}{0.49\hsize}
  \begin{center}
   \includegraphics[height=60mm]{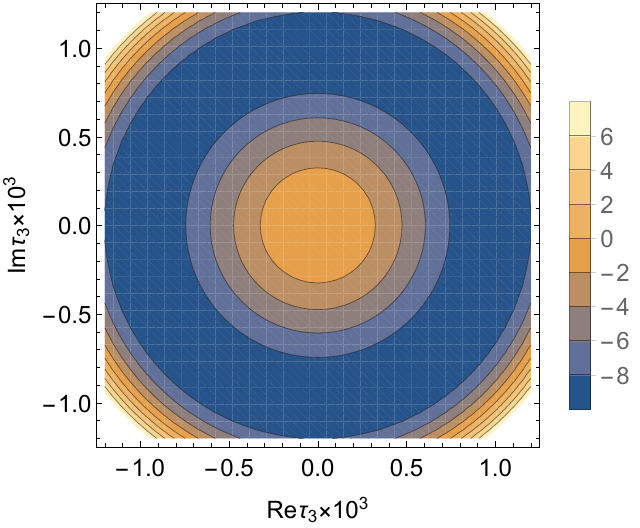}
  \end{center}
 \end{minipage}
    \caption{\label{fig-pot2D}
   The normalized scalar potential  
   $\mathcal{N}\times V/(\lambda_X \Lambda_X^4)$
   of $\tau_3$ around the fixed point $(\tau_1,\tau_3) = (\omega,0)$, 
   where the normalization factor is 
   $\mathcal{N} = 10^5$ ($10^{13}$) 
   on the left (right) panel. 
   The relative size of the CW potential 
   is chosen to 
   $\eps_Q=10^{-2}$~(left)
   and 
   $\eps_Q=10^{-6}$~(right). 
   }
\end{figure}

\begin{figure}[t]
\begin{minipage}{0.49\hsize}
  \begin{center}
  \includegraphics[height=65mm]{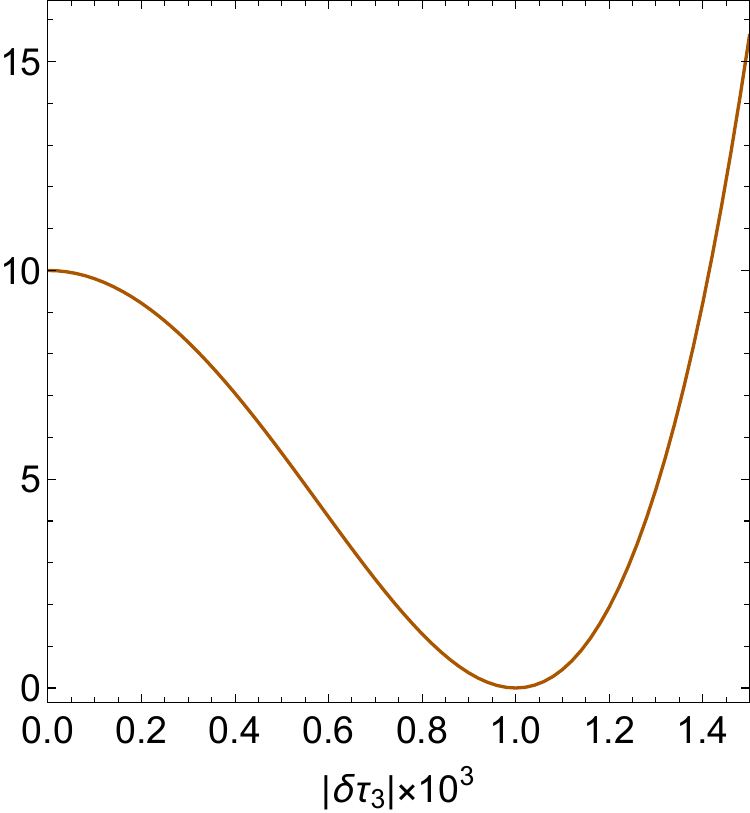}
  \end{center}
 \end{minipage}
 \begin{minipage}{0.49\hsize}
  \begin{center}
   \includegraphics[height=65mm]{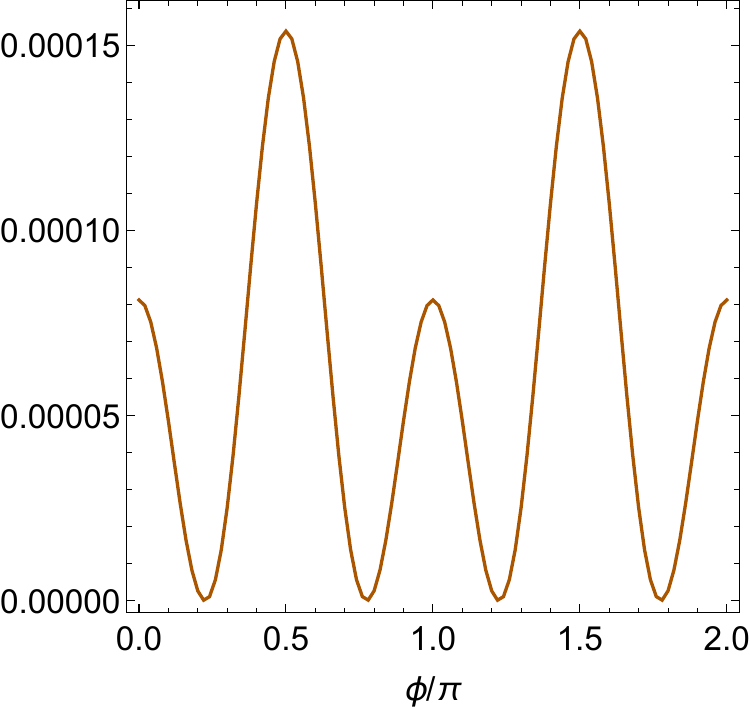}
  \end{center}
 \end{minipage}
    \caption{\label{fig-pot1D}
    The normalized scalar potential 
    along $\abs{\dta_3}$ (left) and $\phi/\pi$ (right), 
    where $\mathcal{N} = 10^{13}$ and $\eps_Q = 10^{-6}$. 
    Here, zero of the potential is set at the minimum, 
    or equivalently we plot $V+\vev{V}$ before the normalization. 
    }
\end{figure}

Fig.~\ref{fig-pot2D} shows the normalized scalar potential  
$\mathcal{N}\times V/(\lambda_X \Lambda_X^4)$ of $\tau_3$ 
around the fixed point $(\tau_1,\tau_3) = (\omega,0)$, 
where the normalization factor is 
$\mathcal{N} = 10^5$ ($10^{13}$) on the left (right) panel. 
On the left panel, 
the relative size of the CW potential is chosen to $\eps_Q=10^{-2}$, 
so that $\abs{\dta_3} \simeq 0.1$. 
Since the deviation is not so small, 
the shape of the potential visibly depends on the phase direction 
around the minimum, 
whereas it looks phase independent near the center $\tau_3 \sim 0$. 
The right panel shows the case of $\eps_Q = 10^{-6}$.  
In this case, the minimum is at $\abs{\delta \tau_3} \simeq 10^{-3}$, 
and the potential is approximately phase independent at the minimum.

We can see this more explicitly in Fig.~\ref{fig-pot1D}
which shows the same potential as the right panel of Fig.~\ref{fig-pot2D}. 
The left panel shows the potential 
along the radial direction with $\phi \simeq 0.22\pi$ 
where is the minimum of the phase direction.  
The minimum resides at $\abs{\dta_3} \simeq 10^{-3}$. 
The potential for the angular direction $\phi$ is shown in the right panel. 
We see that the potential height is much smaller 
than that for the radial direction by $\order{\eps_Q}$. 
This potential has four degenerate minimum 
at $\phi/\pi \sim 0.22,0.78,1.22,1.78$, 
which are not on the CP conserving points 
$\phi \equiv 0$ mod $\pi/2$. 
Thus this potential breaks CP spontaneously.

We discuss the masses of the modulus $\tau_3$. 
The kinetic term of the modulus before canonical normalization is given by 
\begin{align}
\mathcal{L}_{\mathrm{kin}}  
 =  \frac{h}{2} \frac{(\ita_1)^2 + (\ita_3)^2}
                     {\left( (\ita_1)^2 - (\ita_3)^2 \right)^2} 
                \abs{\partial \dta_3}^2.  
\end{align}
At the fixed point $(\tau_1, \tau_3) = (\omega,0)$, 
the canonically normalized fields are given by 
\begin{align}
    \sigma := \sqrt{f_3} \abs{\dta_3}, 
    \quad 
    a := \sqrt{f_3} \vev{\abs{\dta_3}} \phi,
\end{align}
where 
\begin{align}
    f_3 := M_p \sqrt{\frac{h \left\{(\ita_1)^2 + (\ita_3)^2\right\}}
                     {\left\{ (\ita_1)^2 - (\ita_3)^2 \right\}^2} }
         \simeq M_p \sqrt{\frac{4h}{3}}.
\end{align}
The masses squared of the canonically normalized fields at the leading order in $\eps_Q$ are given by 
\begin{align}
    m_\sigma^2 = \frac{2\lambda_X\Lambda_X^2}{f_3^2}, 
\quad 
m_a^2 = 
\eps_Q^2 
(-f_X^{\prime\prime}(\phi) 
+ 2f_Q^{\prime\prime}(\phi)) 
\frac{m_\sigma^2}{2}, 
\end{align}
thus the mass of the phase mode $a$, named axion hereafter, 
is lighter than that of the radial mode $\sigma$ by $\eps_Q$.

The existence of the light axion is interesting 
for phenomenology.  
It could resolve the strong CP problem~\cite{Higaki:2024jdk}, 
and would contribute to the dark matter 
and/or dark radiation~\cite{Jung:2024bgi} 
as in the $SL(2,\mathbb{Z})$ case. 
Since we are utilizing the trivial singlet modular forms 
in the stabilization, 
interpretation of the axion mode in this scenario is not clear. 
In Ref.~\cite{Higaki:2024jdk}, 
the symmetry associated with the $T$ transformation $\tau \to \tau+1$ 
in $SL(2,\mathbb{Z})$ can be interpreted as the accidental $U(1)_{\mathrm{PQ}}$ 
for the QCD axion since a non-trivial singlet modular form 
under the $T$ transformation is coupled to a vector-like quarks. 
Such direct interpretation is absent in the current case, 
since the modular forms are trivial singlets. 
It is interesting to study its interpretation and  phenomenological consequences, 
but these are beyond the scope of this paper.

\subsection{Two Siegel modular forms}
\label{sec:Sp4Zmodel2}

We introduce two matter superfields $X_\alpha$ 
with the modular weight $k_{X_\alpha}$, where $\alpha = 1,2$ 
to stabilize both of the moduli $\tau_1$ and $\tau_3$. 
The K\"ahler potential and the superpotential are given by 
\begin{align}
    K &= -h \ln \left( \det (-i\tau + i \Bar{\tau})\right) 
    + \sum_{\alpha=1}^2 
    \frac{|X_\alpha|^2}{\det (-i\tau + i \Bar{\tau})^{k_{X_\alpha}}}\,,
    \nonumber\\
    W &= \sum_\alpha \Lambda_{X_\alpha}^2
    Y_{\mathbf{1}_\alpha}^{(k_{{\alpha}})} X_\alpha,
\end{align}
where $\Lambda_{X_\alpha}$ is a mass scale, 
$\Ykr{k_\alpha}{\mathbf{1}_\alpha}$ is a modular form with representation $1_\alpha$ 
and  $k_\alpha$ is a modular weight with $\alpha=1,2$. 
The weight satisfies $k_\alpha = k_{X_\alpha} - h$ for the modular invariance. 
Note that we can stabilize the modulus $(\tau_1,\tau_3) \simeq (\omega,0)$ 
by choosing $(k_1, k_2) = (2,6)$ and $(\mathbf{1}, \mathbf{1_c})$ 
as in the radiative stabilization case 
together with the hierarchy $\Lambda_{X_1} \gg \Lambda_{X_2}$, 
since the CW potential is mostly the quadratic potential $\sim \abs{\Ykr{6}{1_c}}^2$
which is the same as that from the matter coupling to $X_2$. 
Such a scale hierarchy would be naturally realized 
in the radiative stabilization case due to $ \eps_Q \sim m_0 m_Q/\Lambda_X^2 \ll 1$.  
We do not repeat the analysis for such a case, 
rather we focus on the stabilization away from the fixed point $(\omega,0)$.

The supersymmetric condition $D_IW =0$ with 
$I= X_1, X_2, \tau_1, \tau_3$ has a solution at which 
\begin{align}
    \langle Y_{\mathbf{1}}^{(k_{\alpha})} (\tau)\rangle 
    = \langle X_\alpha \rangle = 0.
\label{eq:vev_sp4z2}
\end{align}
From the expansion of the superpotential in Eq.~\eqref{eq:vev_sp4z2} 
up to the second order of 
$\delta X_\alpha = X_\alpha- \langle X_\alpha\rangle$ and 
$\delta \tau_{i} = \tau_{i} - \langle\tau_{i}\rangle$ with $i=1,3$:
\begin{align}
    W = \sum_{\alpha=1}^2 \Lambda_{X_\alpha}^2 \delta X_\alpha \left( 
    \frac{\partial Y_{\mathbf{1}_\alpha}^{(k_\alpha)}}{\partial \tau_1} \delta \tau_1 
    +
    \frac{\partial Y_{\mathbf{1}_\alpha}^{(k_{\alpha})}}{\partial \tau_3}\delta \tau_3 \right),
\label{eq:W_two}
\end{align}
both moduli fields $\tau_1$ and $\tau_3$ have non-zero masses 
when two directions 
$\partial_{\tau_1} \Ykr{k_\alpha}{\mathbf{1}_\alpha} \delta \tau_1 
+\partial_{\tau_3} \Ykr{k_\alpha}{\mathbf{1}_\alpha} \delta \tau_3$ 
for $\alpha=1,2$ are not parallel to each other. 
For instance, the singlet modular form of weight four 
is a square of that of weight two, i.e. 
$Y_{\mathbf{1}_a}^{(4)}= (Y_{\mathbf{1}}^{(2)})^2$, 
so the ratio of the derivatives is unity 
and hence there remains a massless direction. 
For more details about the modular forms, see Appendix~\ref{app:sp4}.

\begin{figure}[t]
\begin{minipage}{0.49\hsize}
  \begin{center}
  \includegraphics[height=50mm]{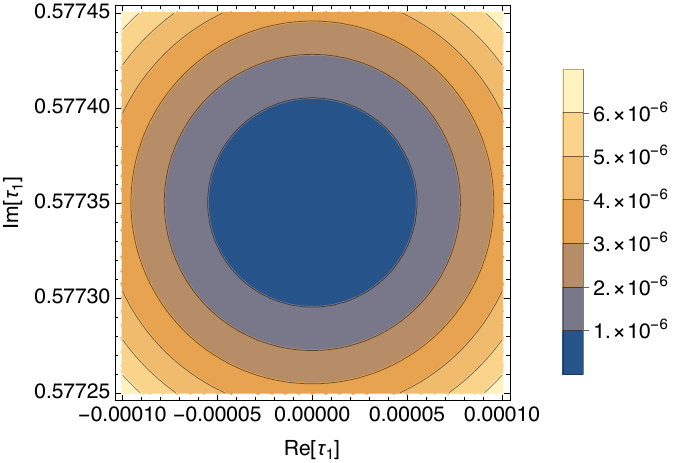}
  \end{center}
 \end{minipage}
 \begin{minipage}{0.49\hsize}
  \begin{center}
   \includegraphics[height=50mm]{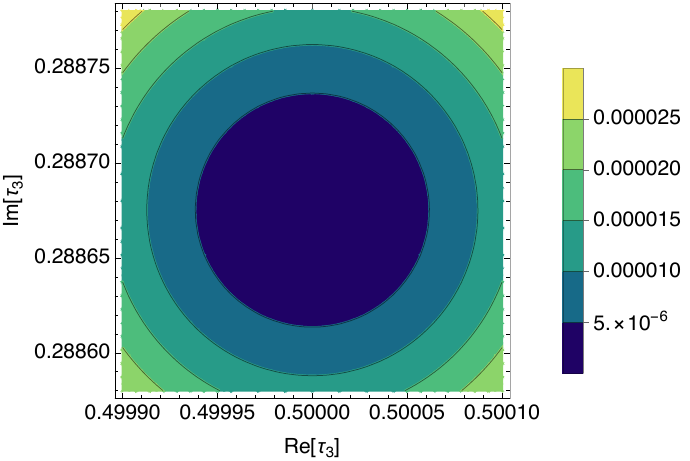}
  \end{center}
 \end{minipage}
    \caption{\label{fig-twoSiegels}
    Scalar potential in units of $\Lambda_{X_1}=\Lambda_{X_2}$ as a function of $\tau_1$ and $\tau_3$ at $\langle Y_{\mathbf{1}}^{(2)} (\tau)\rangle = \langle Y_{\mathbf{1}_b}^{(4)} (\tau)\rangle  = 0$, where we set $\tau_3=\langle\tau_3\rangle$ and $\tau_1 =\langle \tau_1 \rangle$ in Eq. \eqref{eq:Y2Y4b}} in the left and right figures, respectively.
\end{figure}

For the singlet modular forms up to the weight six, 
there are three combinations which can satisfy Eqs.~\eqref{eq:vev_sp4z2} and \eqref{eq:W_two}, 
namely $(\Ykr{2}{1}, \Ykr{4}{1_b})$, $(\Ykr{2}{1}, \Ykr{6}{1_c})$
and $(\Ykr{4}{1_b}, \Ykr{6}{1_c})$. 
For the first combination, both of the modular forms are zero, 
but we cannot numerically find solutions leading to two massive directions in later two cases. 
In the latter two cases, the first derivative of modular forms with respect to moduli fields 
are not independent of each other, 
and there still exists a massless direction 
along which CP-conserving and -breaking vacuum is degenerate~\footnote{Such a coexistence of CP-conserving and -breaking vacuum can also be seen in other types of moduli stabilization in the framework of modular symmetric models \cite{Kobayashi:2020hoc}, which is applied to realize the spontaneous CP violation in \cite{Higaki:2024pql}.}. 
In the first case,  
$\langle Y_{\mathbf{1}}^{(2)} (\tau)\rangle 
= \langle Y_{\mathbf{1}_b}^{(4)} (\tau)\rangle  = 0$, 
We find the minimum of the potential is at 
    \begin{align}
    \langle \tau_1 \rangle \simeq 0.57735i,\qquad
    \langle \tau_3 \rangle \simeq 0.50000+0.28867i.
\label{eq:Y2Y4b}
\end{align}
The shape of the potential around this point 
is shown in Fig.~\ref{fig-twoSiegels}. 
The left panel is the potential on $\tau_1$ plane with $\tau_3 = \vev{\tau_3}$ 
and the right panel is that on $\tau_3$ plane 
with $\tau_1 = \vev{\tau_1}$. 
We see that both $\tau_1$ and $\tau_3$ 
reside at the minimum of the potential. 
Moreover, the Hessian matrix at the minimum is given by 
\small
\begin{align}
    \left(\begin{array}{cccc}
    245.79  & -245.829-0.077639i & 0 & 0\\
     -245.829+0.077639i   & 245.833 & 0 & 0\\
    0 & 0 & 245.79  & -245.829-0.077639i\\
     0 & 0 & -245.829+0.077639i   & 245.833\\
  \end{array}\right)\,,
\end{align}
\normalsize
in units of $\Lambda_{X_1}=\Lambda_{X_2}$. Here, we evaluate the Hessian matrix in the basis of canonically normalized fields $
    \{ \Phi_1, \Phi_3, \bar{\Phi}_1, \bar{\Phi}_3\}$
with
\begin{align}
    \begin{pmatrix}
        \Phi_1 \\ \Phi_3 
    \end{pmatrix}
    := \vev{\frac{1}{2\{(\ita_1)^2-(\ita_3)^2\}}
    \begin{pmatrix}
    \mathrm{Im}(\tau_1+\tau_3) &   -\mathrm{Im}(\tau_1+\tau_3) \\ 
     \mathrm{Im}(\tau_1-\tau_3)&  \mathrm{Im}(\tau_1-\tau_3)
    \end{pmatrix}
    }
    \begin{pmatrix}
        \tau_1 \\ \tau_3 
    \end{pmatrix}
    , 
\end{align}
Hence, there is no massless modes 
unlike the potential alone by $\Ykr{2}{1}$.  
We also note that the point $(\omega,0)$ 
satisfy Eq.~\eqref{eq:vev_sp4z2}, but $\tau_3$ remains massless.

The minimum is away from the fixed points 
in the Siegel upper half plane ${\cal H}_2$, but $\CP$ is not broken.
Our numerical analysis reveals that 
the vacuum expectation values of moduli fields in Eq.~\eqref{eq:Y2Y4b} lead to the positive mass squared of all the moduli. 
However, the CP symmetry is still preserved in this vacuum.

\section{Conclusion}
\label{sec:con}

In this paper, we discussed the stabilization of moduli fields by Siegel modular forms 
in the $Sp(2g, \mathbb{{Z}})$ modular invariant theories. 
For a generic modular- and CP-invariant scalar potential, 
we derive the stationary conditions at fixed points under the modular and CP symmetries. 
We demonstrated that the CP-conserving fixed points of $Sp(4,\mathbb{Z})$ 
are indeed stationary points. 
The obtained conditions Eqs.~\eqref{eq:Vtauij_general} and \eqref{eq:bartauij_general_sij} 
can be applicable to the fixed points 
in a generic $Sp(2g, \mathbb{Z})$ modular symmetry.

For concreteness, 
we proposed the stabilization mechanism of multi two moduli fields by utilizing 
Siegel modular forms under $S^4 \times \mathbb{Z}_2$ of $Sp(4,\mathbb{Z})$ in Sec. \ref{sec:3} 
and $\tilde{\Delta}(96)$ of $Sp(6,\mathbb{Z})$ in Appendix \ref{app:sp6}. 
By introducing matter fields in a modular-invariant way, 
we obtain the modular-invariant scalar potential. 
It turns out that the stabilization of all the moduli fields requires the deviation from fixed points of $\tau$ in the Siegel upper half plane. 
The coexistence of two Siegel modular forms leads to the stabilization of all the moduli fields in a supersymmetric Minkowski minimum at the classical level, but it requires a specific choice of Siegel modular forms. 
When some of the moduli fields are not stabilized at the classical level,  quantum corrections from matter fields 
may stabilize remaining moduli fields.

We demonstrated a stabilization scenario 
in which the tree-level potential is stabilized 
at one of the fixed points, $(\tau_1, \tau_3) = (\omega,0)$ 
with the massless mode $\sim \tau_3$, 
and then the radiative correction generates the mass of $\tau_3$. 
The radiative potential deviates the location of the minimum 
due to its negative mass squared at the fixed point. 
Interestingly, the phase direction is approximately massless 
similarly to the Nambu-Goldstone mode of the Higgs potential in the SM, 
but with slight phase dependence at $\order{\abs{\dta_3}^6}$. 
There will be phenomenological applications of 
the light mode, such as identifying it as axion (like particle). 
The stabilization nearby a fixed point will be interesting to explain 
the hierarchical structure of the quarks and leptons in the SM 
utilizing the residual symmetries as done in the $SL(2,\intZ)$ case as well as $Sp(6,\mathbb{Z})$~\cite{Feruglio:2021dte,Novichkov:2021evw,Petcov:2022fjf,Ishiguro:2022pde,Kikuchi:2023cap,Abe:2023ilq,Kikuchi:2023jap,Abe:2023qmr,Petcov:2023vws,Abe:2023dvr,deMedeirosVarzielas:2023crv,Kikuchi:2023dow,Kikuchi:2023fpl}.  
Those applications of Siegel modular forms to phenomenology are our future works.

\acknowledgments

This work was supported in part JSPS KAKENHI Grant Numbers JP23K03375 (T.K.), JP24KJ0249 (K.N.), and JP23H04512 (H.O).
The work of J.K. was supported by IBS under the project code, IBS-R018-D1.

\appendix

\section{Siegel modular forms of $Sp(4,\mathbb{Z})$}
\label{app:sp4}

It was known that the space of the Siegel modular forms of weight two under $Sp(4,\mathbb{Z})$, ${\cal M}(\Gamma_2(2))$, 
is spanned by five polynomials:
\begin{align}
    p_0 &= \Theta[00]^4(\tau) + \Theta[01]^4(\tau) +\Theta[10]^4(\tau) +\Theta[11]^4(\tau)\,,
    \nonumber\\
    p_1 &= 2\left(\Theta[00]^2(\tau)\Theta[01]^2(\tau) +\Theta[10]^2(\tau)\Theta[11]^2(\tau)\right)\,,
    \nonumber\\
    p_2 &= 2\left(\Theta[00]^2(\tau)\Theta[10]^2(\tau) +\Theta[01]^2(\tau)\Theta[11]^2(\tau)\right)\,,
    \nonumber\\
    p_3 &= 2\left(\Theta[00]^2(\tau)\Theta[11]^2(\tau) +\Theta[01]^2(\tau)\Theta[10]^2(\tau)\right)\,,
    \nonumber\\
    p_4 &= 4\Theta[00](\tau)\Theta[01](\tau)\Theta[10](\tau)\Theta[11](\tau)\,,
\end{align}
where $\Theta[\sigma](\tau)$ is the second order Theta constant:
\begin{align}
    \Theta[\sigma](\tau) = \sum_{m\in \mathbb{Z}^2} e^{2\pi i (m+\sigma/2)\tau(m+\sigma/2)^T}\,.
\label{eq:thetaconstant}
\end{align}
The finite modular group $S_4\times \mathbb{Z}_2$ can be generated by these elements $\{\mathcal{S},\mathcal{T},\mathcal{V}\}$ 
satisfying 
\begin{align}
    \mathcal{S}^2=\mathcal{T}^3=(\mathcal{S}\mathcal{T})^4=1,\quad 
    \mathcal{V}^2 =1,\quad \mathcal{S}\mathcal{V}=\mathcal{V}\mathcal{S},\quad\mathcal{T}\mathcal{V}=\mathcal{V}\mathcal{T}.
\end{align}
Their explicit forms are given by
\begin{align}
    \mathcal{S}=G_1,\quad \mathcal{T}=(G_3 G_2)^4,\quad \mathcal{V}=(G_3 G_2)^3, 
\end{align}
with
\begin{align}
  G_1=T_1T_2=\left(\begin{array}{cccc}
    1   & 0&1&0 \\
    0   & 1&0&1\\
    0&0&1&0\\
    0&0&0&1\\
  \end{array}\right),\quad
  G_2=T_3=\left(\begin{array}{cccc}
   1    &  0&0&1\\
     0  & 1&1&0\\
     0&0&1&0\\
     0&0&0&1\\
  \end{array}\right), 
\\ \notag 
  \end{align}
  \begin{align}
  G_3=S=\left(\begin{array}{cccc}
     0  &  0&1&0\\
    0   & 0&0&1\\
    -1&0&0&0\\
    0&-1&0&0\\
  \end{array}\right),\quad
  G_4=\left(\begin{array}{cccc}
   1    & 0&0&0 \\
    0   & -1&0&0\\
    0&0&1&0\\
    0&0&0&-1\\
  \end{array}\right).
\end{align}

When we restrict ourselves to $\tau_1=\tau_2$ in the Siegel upper half plane, 
the space of the Siegel modular forms is spanned by four polynomials $\{p_0, p_1, p_3, p_4\}$ with $p_2=p_3$. In the following, we list the singlet modular forms of weight two, four and six under $S_4\times \mathbb{Z}_2$: 
\paragraph{Weight 2}
\begin{align}
Y_{\mathbf{1}}^{(2)}(\tau)&=
p_0 (\tau) +3p_3 (\tau) =:Y_4(\tau)
.
\label{eq:weight2S4Z2}
\end{align}
\paragraph{Weight 4}
\begin{align}
Y_{\mathbf{1}}^{(4_a)}(\tau)&= Y_4^2,
\nonumber\\
Y_{\mathbf{1}}^{(4_b)}(\tau)&= Y_1^2 +2 Y_2 Y_3,    
\end{align}
with
\begin{align}
    Y_1(\tau) &\equiv p_0(\tau)+4p_1(\tau)-p_3(\tau),
    \nonumber\\
    Y_2(\tau) &\equiv p_0(\tau)-2p_1(\tau)-p_3(\tau)-2i\sqrt{3}p_4,
    \nonumber\\
    Y_3(\tau) &\equiv p_0(\tau)-2p_1(\tau)-p_3(\tau)+2i\sqrt{3}p_4.
\end{align}
\paragraph{Weight 6}
\begin{align}
Y_{\mathbf{1}}^{(6_a)}(\tau)&= Y_4^3,
\nonumber\\
Y_{\mathbf{1}}^{(6_b)}(\tau)&= Y_4(Y_1^2 +2 Y_2 Y_3),    
\nonumber\\
Y_{\mathbf{1}}^{(6_c)}(\tau)&= 2(Y_1^3 + Y_2^3 + Y_3^3 -3 Y_1 Y_2 Y_3).    
\end{align}

\begin{table}[t] 
\centering
\caption{Values of singlet modular forms up to weight six at the fixed point of $\Gamma_2 = Sp(4,\mathbb{Z})$ in the Siegel upper half plane ${\cal H}_2$. Note that the values of other modular forms can be evaluated from $Y_{\mathbf{1}}^{(2)}(\tau)$, $Y_{\mathbf{1}}^{(4_b)}(\tau)$ and $Y_{\mathbf{1}}^{(6_c)}(\tau)$.}
\label{tab:Yzero_sp4}
\begin{tabular}{c|ccc} \hline
  Fixed points & $Y_{\mathbf{1}}^{(2)}(\tau)$ & $Y_{\mathbf{1}}^{(4_b)}(\tau)$ & $Y_{\mathbf{1}}^{(6_c)}(\tau)$ 
  \\ \hline\hline 
$
\begin{pmatrix}
  \zeta & \zeta + \zeta^{-2}  \\
  \zeta + \zeta^{-2}   & -\zeta^{-1}
\end{pmatrix}
$  &
0.366-0.302$i$ &
3.01-2.92$i$ &
-13.7-20.4$i$ 
\\ \hline
$
\begin{pmatrix}
  \eta & \frac{1}{2}(\eta -1)  \\
  \frac{1}{2}(\eta -1) & \eta
\end{pmatrix}
$ &
0.207-0.584$i$ &
6.278+5.072$i$ &
-21.9+13.45$i$
\\ \hline
$
\begin{pmatrix}
  i & 0  \\
  0 & i
\end{pmatrix}
$ &
1.46 &
6.36 &
12.3 
\\ \hline
$
\begin{pmatrix}
  \omega & 0  \\
  0 & \omega
\end{pmatrix}
$ &
0 &
0 &
-33.2 \\
\hline
$\frac{i}{\sqrt{3}}
\begin{pmatrix}
  2 & 1  \\
  1 & 2
\end{pmatrix}
$ &
1.81 &
4.00 &
3.50 \\
\hline
$\begin{pmatrix}
  \omega & 0  \\
  0 & i
\end{pmatrix}$
& 
1.21-0.270$i$ &
6.84-0.113$i$ &
14.2+2.42$i$ 
\\ \hline
\end{tabular}
\end{table}

As commented in Sec.~\ref{sec:Sp4Zmodel1}, 
the singlet modular forms up to weight six are zero at the fixed point \eqref{eq:vev_sp4z_24a1} 
except $\Ykr{6}{1_c}$. 
Indeed, we can show this as follows. 
Let us consider the modular transformation of the singlet modular form 
$Y_{\mathbf{1}}^{(k_Y)}$:
\begin{align}
    Y_{\mathbf{1}}^{(k_Y)}(h_\ast \tau) = \bigl[{\rm det}(C \tau +D)]^{k_Y} Y_{\mathbf{1}}^{(k_Y)}(\tau),
\end{align}
with
\begin{align}
    h_\ast =
    \begin{pmatrix}
        A & B\\
        C & D
    \end{pmatrix}
    =
    \left(
\begin{smallmatrix}
  0 & 0 & 1 & 0  \\
  0 & 0 & 0 & -1  \\
  -1 & 0 & -1 & 0  \\
  0 & 1 & 0 & 1 \\
\end{smallmatrix}
\right), 
\end{align}
being one of generators of the stabilizer group at the fixed point 
$\tau_{\rm fix}= {\rm diag}(\omega, \omega)$ in Table~\ref{tab:Sp4Zfp}. 
Then, \eqref{eq:Ymodulartrf} with $k_Y=2,4,6$ obeys
\begin{align}
    Y_{\mathbf{1}}^{(2)}(\tau_{\rm fix}) &= (\omega +1)^{4}Y_{\mathbf{1}}^{(2)}(\tau_{\rm fix}),
\nonumber\\
    Y_{\mathbf{1}}^{(4)}(\tau_{\rm fix}) &=\omega Y_{\mathbf{1}}^{(4)}(\tau_{\rm fix}),
\nonumber\\
    Y_{\mathbf{1}}^{(6)}(\tau_{\rm fix}) &=Y_{\mathbf{1}}^{(6)}(\tau_{\rm fix}),
\end{align}
which results in $Y_{\mathbf{1}}^{(2)}(\tau_{\rm fix})=Y_{\mathbf{1}}^{(4)}(\tau_{\rm fix})=0$, 
i.e., $Y_{4}(\tau_{\rm fix}) = Y_{1}(\tau_{\rm fix})^2 + 2Y_2(\tau_{\rm fix}) Y_3(\tau_{\rm fix}) = 0$. 
Hence, the modular forms of higher weight constructed by $Y_{\mathbf{1}}^{(2)}$ and $Y_{\mathbf{1}}^{(4_b)}$ are also zero at the fixed point $\tau=\tau_{\rm fix}$. 
Specifically, only $Y_{\mathbf{1}}^{(6_d)}$ is nonzero at $\tau=\tau_{\rm fix}$ among the singlet modular forms up to the weight six. We have focused on the specific fixed point, but we check all the fixed points of $\tau$ whether the singlet modular forms are zero, 
as shown in Table~\ref{tab:Yzero_sp4}.

We can also analytically show that the first derivative of $Y_{\mathbf{1}}^{(k_Y)}$ with respect to $\tau_3$ vanishes at $\tau_3 = 0$ which can be 
derived from the properties of the second order Theta constant \eqref{eq:thetaconstant}:
\begin{align}
    \frac{d}{d\tau_3} \Theta[00](\tau)\biggl|_{\tau=\tau_{\rm fix}} &= \frac{d}{d\tau_3} \sum_{m_1,m_2} e^{2\pi i \{(m_1^2+m_2^2)\tau_1 + 2m_1 m_2 \tau_3 \}}\biggl|_{\tau=\tau_{\rm fix}}
    \nonumber\\
    &= \sum_{m_1,m_2} 2m_1m_2 e^{2\pi i \{(m_1^2+m_2^2)\tau_1 + 2m_1 m_2 \tau_3 \}}\biggl|_{\tau=\tau_{\rm fix}}
    \nonumber\\
    &= \sum_{m_1,m_2} 2m_1m_2\,e^{2\pi i (m_1^2+m_2^2)\tau_1}\biggl|_{\tau=\tau_{\rm fix}}
    =0.
\end{align}
Similarly, one can show that the first derivatives of the other second order Theta constants $\Theta[01](\tau)$ and $\Theta[11](\tau)$ with respect to $\tau_3$ are zero at the fixed point \eqref{eq:vev_sp4z_24a1} by the same reason as in $\Theta[00](\tau)$, i.e.,
\begin{align}
    \frac{d}{d\tau_3} \Theta[01](\tau)\biggl|_{\tau=\tau_{\rm fix}} =0,
    \qquad
    \frac{d}{d\tau_3} \Theta[11](\tau)\biggl|_{\tau=\tau_{\rm fix}} =0,
\end{align}
with
\begin{align}
    \Theta[01](\tau) &= \sum_{m_1, m_2} e^{2\pi i \bigl[ \left\{\left(m_1+\frac{1}{2}\right)^2+m_2^2 \right\} \tau_1 + 2\left(m_1+\frac{1}{2}\right) m_2 \tau_3  \bigl]}\,,
    \nonumber\\
    \Theta[11](\tau) &= \sum_{m_1, m_2} e^{2\pi i \bigl[ \left\{\left(m_1+\frac{1}{2}\right)^2+\left(m_2+\frac{1}{2}\right)^2 \right\} \tau_1 + 2\left(m_1+\frac{1}{2}\right) \left(m_1+\frac{1}{2}\right) \tau_3  \bigl]}\,.    
\end{align}
Since all the singlet modular forms are determined by the second order Theta constants as in \eqref{eq:weight2S4Z2}, it turns out that the first derivative of $Y_{\mathbf{1}}^{(k_Y)}(\tau)$ with respect to $\tau_3$ vanishes at the fixed point \eqref{eq:vev_sp4z_24a1}. 
We can generalize this to odd derivatives of the modular forms, i.e. 
\begin{align}
\left( \frac{\partial}{\partial \tau_3}\right)^{2n+1} \Ykr{k_Y}{1} = 0  
\end{align}
at $\tau_3 = 0$.

\section{$Sp(6,\mathbb{Z})$ modular models}
\label{app:sp6}

In this section, we study the Siegel modular group $\Gamma_3=Sp(6,\mathbb{Z})$ 
and its finite subgroups. 
After reviewing $\Gamma_3$ realized in a specific top-down model building in Sec. \ref{sec:sp6}, 
we discuss the stabilization of the moduli fields by introducing a matter 
superpotential in Sec. \ref{sec:Sp6Zmodel}.

\subsection{Finite Siegel modular group of $Sp(6,Z)$}
\label{sec:sp6}

In this section, we discuss the Siegel modular group $\Gamma_3=Sp(6,\mathbb{Z})$, 
which can be generated by four elements:
\begin{align}
    T_m = \left(\begin{array}{cc}
      \mathbbm1_3 & B_m \\
        0_3 & \mathbbm1_3
    \end{array}\right)\,,
    \quad
    S=\left(
    \begin{array}{cc}
       0_3  &  \mathbbm1_3\\
      -\mathbbm1_3   & 0_3
    \end{array}\right), 
\end{align}
with
\begin{align}
    B_1=\left(\begin{array}{ccc}
      1   & 0  & 0\\
      0   & 0  & 0\\
      0   & 0  & 0
    \end{array}\right)\,,
    \quad
    B_2=\left(\begin{array}{ccc}
      0   & 0  & 0\\
      0   & 1  & 0\\
      0   & 0  & 0
    \end{array}\right)\,,
    \quad
    B_3=\left(\begin{array}{ccc}
      0   & 0  & 0\\
      0   & 0  & 0\\
      0   & 0  & 1
    \end{array}\right)
    ,
    \nonumber\\
    B_4=\left(\begin{array}{ccc}
      0   & 1  & 0\\
      1   & 0  & 0\\
      0   & 0  & 0
    \end{array}\right)\,,
    \quad
    B_5=\left(\begin{array}{ccc}
      0   & 0  & 0\\
      0   & 0  & 1\\
      0   & 1  & 0
    \end{array}\right)\,,
    \quad
     B_6=\left(\begin{array}{ccc}
      0   & 0  & 1\\
      0   & 0  & 0\\
      1   & 0  & 0
    \end{array}\right)
    ,
\end{align}
under which the moduli fields transform as in Eq.~\eqref{eq:moduliTrf}. 
The Siegel modular group allows the existence of finite modular groups. 
In this paper, we focus on the finite Siegel modular group $\tilde{\Delta}(96)\simeq \Delta(48)\rtimes \mathbb{Z}_8$ which is also isomorphic to $\Tilde{\Gamma}_8$ ([768, 1085324] in GAP system), consisted by two generators $\Tilde{S}$ and $\Tilde{T}$~\footnote{Note that this $\Tilde{\Gamma}_8$ modular group is also understood by the finite metaplectic modular group, as realized in magnetized D-brane models~\cite{Almumin:2021fbk,Ishiguro:2023jqb}.}. 
They obey
\begin{align}
    &\widetilde{S}^2 = \Tilde{R},\qquad
    (\widetilde{S}\Tilde{T})^3 = \Tilde{R}^4 = 1,\qquad
    \Tilde{T}\Tilde{R} = \Tilde{R}\Tilde{T},\qquad
    \tilde{T}^{8}=\mathbb{I},
    \nonumber\\
&\widetilde{S}^5\tilde{T}^6\widetilde{S}\tilde{T}^4\widetilde{S}\tilde{T}^2\widetilde{S}\tilde{T}^4 = (\Tilde{S}^{-1}\Tilde{T}^{-1} \Tilde{S} \Tilde{T} )^3 = \mathbb{I}.
\label{eq:meta_generator}
\end{align}

Let us discuss the relation between $\Tilde{\Gamma}_8$ and $\Gamma_3=Sp(6,\mathbb{Z})$ modular groups. 
To obtain the modular forms, we refer to the derivation based on the fermionic zero modes in ten-dimensional non-Abelian supersymmetric Yang-Mills theory on $T^{6}$ with $U(1)$ magnetic fluxes~\cite{Kikuchi:2023dow}. 
The $\Gamma_3$ can be regarded as a geometric symmetry of six-dimensional torus. 
By solving the Dirac equation for ten-dimensional gauginos, one can obtain the wavefunction of massless zero modes. 
It was known that when we consider $\mathbb{Z}_2$-even mode under $\vec{z}\rightarrow - \Vec{z}$, the solution is given by
\begin{align}
\begin{pmatrix}
\zeta(\tau) \\
\eta(\tau) \\
\theta(\tau) \\
\end{pmatrix}
\equiv
\begin{pmatrix}
\sqrt{2}\psi_N^{\begin{psmallmatrix}0\\-1\\0\\\end{psmallmatrix}}(0,\tau) \\
\psi_N^{\begin{psmallmatrix}0\\0\\0\\\end{psmallmatrix}}(0,\tau) \\
-\psi_N^{\begin{psmallmatrix}0\\0\\-1\\\end{psmallmatrix}}(0,\tau) \\
\end{pmatrix}, 
\label{eq:zeromodes}
\end{align}
where $\psi_N^{\vec{j}}(\vec{z},\tau)$ is given by the so-called Riemann-theta function $\vartheta
\begin{bmatrix}
\vec{j}N^{-1} \\ 0 \\
\end{bmatrix}
(N\vec{z},N\tau)$~\cite{Cremades:2004wa}
\begin{align}
\psi_N^{\vec{j}}(\vec{z},\tau) = {\cal N} e^{i\pi (N\vec{z})^T(\textrm{Im}\tau)^{-1}\textrm{Im}\vec{z}} \cdot \vartheta
\begin{bmatrix}
\vec{j}N^{-1} \\ 0 \\
\end{bmatrix}
(N\vec{z},N\tau).
\end{align}
Here and in what follows, we choose $U(1)$ flux quanta:
\begin{align}
N = 
\begin{pmatrix}
-1&1&-1\\
1&-1&-1\\
-1&-1&2\\
\end{pmatrix}
,
\end{align}
which determines the degeneracy of zero modes labeled by $\vec{j}\in\mathbb{Z}^3$. 
This structure is originated from the periodicity of zero mode wavefunctions. 
The explicit form of the Riemann-theta function is defined as
\begin{align}
\vartheta
\begin{bmatrix}
\vec{a} \\ \vec{b}
\end{bmatrix}(\vec{z},\tau^\prime) =
\sum_{\vec{m}\in\mathbb{Z}^3}
e^{\pi i(\vec{m}+\vec{a})^T\tau^\prime (\vec{m}+\vec{a})} e^{2\pi i(\vec{m}+\vec{a})^T(\vec{z}+\vec{b})},
\end{align}
with $\vec{a},\vec{b}\in\mathbb{R}^3$.
Note that $\tau^\prime=N\tau$ is defined in the Siegel upper half plane ${\cal H}_3$. 
The normalization condition is chosen as 
\begin{align}
\int_{T^6} d^3zd^3\bar{z}\, \psi_N^{\vec{j}} (\psi_N^{\vec{k}})^* = (2^3 \det (\textrm{Im}\tau))^{-1/2} \delta_{\vec{j},\vec{k}}.
\end{align}

To see the  structure of wavefunctions, let us consider the following symplectic transformations:
\begin{align}
    S=\left(
    \begin{array}{cc}
       0  &  \mathbbm1_3\\
      -\mathbbm1_3   & 0
    \end{array}\right)\,,
    \qquad
{\cal T}_m =
\begin{pmatrix}
1_3 & {\cal B}_m \\
0_3 & 1_3 \\
\end{pmatrix}
,
\end{align}
with
\begin{align}
&{\cal B}_1 =
\begin{pmatrix}
-1 & 1 & 0 \\
1 & -1 & 0 \\
0 & 0 & 0 \\
\end{pmatrix}
,
\quad
{\cal B}_{2} = 
\begin{pmatrix}
-2 & 1 & -1 \\
1 & -2 & -1 \\
-1 & -1 & 1 \\
\end{pmatrix}
,\quad
{\cal B}_{3} =
\begin{pmatrix}
-1 & -1 & 0 \\
-1 & -1 & 0 \\
0 & 0 & -2 \\
\end{pmatrix}
,
\end{align}
under which the wavefunctions \eqref{eq:zeromodes} transform as
\begin{align}
\begin{pmatrix}
\zeta \\ \eta \\ \theta \\
\end{pmatrix}
\xrightarrow{S}
\sqrt{\det (-\Omega)} \rho(S)
\begin{pmatrix}
\zeta \\ \eta \\ \theta \\
\end{pmatrix}, \quad
\begin{pmatrix}
\zeta \\ \eta \\ \theta \\
\end{pmatrix}
\xrightarrow{{\cal T}_i}
\rho({\cal T}_i)
\begin{pmatrix}
\zeta \\ \eta \\ \theta \\
\end{pmatrix},
\end{align}
with
\begin{align}
\begin{aligned}
&\rho(S) = e^{-7\pi i/4}
\begin{pmatrix}
0 & \frac{1}{\sqrt{2}}i & \frac{1}{\sqrt{2}}i \\
\frac{1}{\sqrt{2}}i & \frac{1}{2}i & -\frac{1}{2}i \\
\frac{1}{\sqrt{2}}i & -\frac{1}{2}i & \frac{1}{2}i \\
\end{pmatrix}, \quad
\rho({\cal T}_1) =
\begin{pmatrix}
i & 0 & 0 \\
0 & 1 & 0 \\
0 & 0 & 1 \\
\end{pmatrix}, \\
&\rho({\cal T}_2) =
\begin{pmatrix}
e^{7\pi i/4} & 0 & 0 \\
0 & 1 & 0 \\
0 & 0 & -1 \\
\end{pmatrix}, \quad
\rho({\cal T}_3) =
\begin{pmatrix}
-1 & 0 & 0 \\
0 & 1 & 0 \\
0 & 0 & 1 \\
\end{pmatrix}.
\end{aligned} \label{eq:rhos}
\end{align}
From the $S$-transformation, one can extract the value of modular weight as 1/2.
Furthermore, $\rho(S)$ and $\rho({\cal T}_2)$ form the generators of unitary representation under $ \widetilde{\Delta}(96)\simeq \Delta(48)\rtimes Z_8$. 
It means that wavefunctions \eqref{eq:zeromodes} belong to a triplet representation of $\widetilde{\Delta}(96)$ which is described by the Siegel modular forms of weight 1/2 under $\widetilde{\Delta}(96)$. 
Note that such a finite modular symmetry can be discussed in a generic point of the moduli space of $\tau$:
\begin{align}
    \tau=\left(\begin{array}{ccc}
      \tau_1   &  \tau_4 & \tau_5\\
      \tau_4   &  \tau_2 & \tau_6\\
      \tau_5   &  \tau_6 & \tau_3\\
    \end{array}\right)
    \,.
\end{align}

Except for the large complex structure points, there are three fixed points under $\tilde{\Delta}(96)$, as shown in Table \ref{tab:Sp6Zfp}.
\begin{table}[H]
    \centering
    \begin{tabular}{|c|c|}\hline
      Fixed points   &  Generators of the stabilizer group\\
      \hline
$\frac{i}{\sqrt{3}}\left(
\begin{array}{ccc}
  \frac{-1-\sqrt{3}}{2} & \frac{-1+\sqrt{3}}{2} & -1  \\
  \frac{-1+\sqrt{3}}{2} & \frac{-1-\sqrt{3}}{2} & -1  \\
  -1 & -1 & 1  \\
\end{array}
\right)$
& 
$S=\left(
    \begin{array}{cc}
       0  &  \mathbbm1_3\\
      -\mathbbm1_3   & 0
    \end{array}\right)$
\\
\hline
$\frac{i\omega^2}{\sqrt{3}}
\left(
\begin{array}{ccc}
  -\frac{1+i 3\sqrt{3}}{4} & \frac{1}{2}\omega &  \omega^2 \\
  \frac{1}{2}\omega & -\frac{1+i3\sqrt{3}}{4} &  \omega^2 \\
  \omega^2 & \omega^2 &  \omega \\
\end{array}
\right)$
& 
$S{\cal T}_1^{-1}{\cal T}_2=\left(
    \begin{array}{cc}
       0  &  \mathbbm1_3\\
      -\mathbbm1_3   & {\cal C}_1
    \end{array}\right)$
with 
${\cal C}_1=\left(
\begin{array}{ccc}
  1 & 0 & 1   \\
  0 & 1 & 1  \\
  1 & 1 & -1  \\
\end{array}
\right)$
\\
\hline
$\frac{i\omega^2}{\sqrt{3}}
\left(
\begin{array}{ccc}
  -\omega & -1 &  \omega^2 \\
  -1 & -\omega &  \omega^2 \\
  \omega^2 & \omega^2 &  \omega \\
\end{array}
\right)$
& 
$S{\cal T}_1^{-2}{\cal T}_2=\left(
    \begin{array}{cc}
       0  &  \mathbbm1_3\\
      -\mathbbm1_3   & {\cal C}_2
    \end{array}\right)$
with 
${\cal C}_2=\left(
\begin{array}{ccc}
  0 & 1 & 1   \\
  1 & 0 & 1  \\
  1 & 1 & -1  \\
\end{array}
\right)$
\\
\hline
    \end{tabular}
    \caption{Fixed points of $\Tilde{\Delta}(96)$ except for the large complex structure points. Here, we use $\omega = e^{2\pi i/3}$.}
    \label{tab:Sp6Zfp}
\end{table}

From the tensor product of the Siegel modular forms of weight 1/2, one can construct Siegel modular forms of higher weight. 
In the following analysis of the moduli stabilization, we focus on a singlet modular form, in particular, the Siegel modular form of weight 4 under $\widetilde{\Delta}(96)$.

\subsection{Moduli stabilization}
\label{sec:Sp6Zmodel}

In this section, we study the supersymmetric effective action in the framework of $Sp(6,\mathbb{Z})$ modular symmetry. 
As discussed in Sec. \ref{sec:Sp4Zmodel1}, we introduce a chiral superfield $X$ whose effective K\"ahler potential and superpotential are described by
\begin{align}
    K &= -h \ln \left( \det (-i\tau + i \Bar{\tau})\right) + \frac{|X|^2}{\det (-i\tau + i \Bar{\tau})^k}\,,
    \nonumber\\
    W &= Y_{\mathbf{1}}^{(4_a)}(\tau) X,
\label{eq:KWsp6sim}
\end{align}
where $h$ is a positive integer, and $Y_{\mathbf{1}}^{(4_a)}(\tau)$ is one of a singlet modular form with weight 4 under $\widetilde{\Delta}(96)$. 
The explicit form is given in Appendix \ref{app:sp6modular}.

By solving the supersymmetric conditions:
\begin{align}
    \partial_X W &= Y_{\mathbf{1}}^{(4_a)}(\tau) = 0,
    \nonumber\\
    \partial_{\tau_n} W &= (\partial_{\tau_n}Y_{\mathbf{1}}^{(4)}) X = 0,
    \nonumber\\
    W & = 0,
\end{align}
the matter $X$ and moduli $\tau_{n}$ with $n=1,2,3,4,5,6$ can be stabilized at
\begin{align}
    \langle Y_{\mathbf{1}}^{(4_a)}(\tau)\rangle = \langle X \rangle = 0.
\end{align}
As discussed in the $Sp(4,\mathbb{Z})$ case, the supersymmetric conditions \eqref{eq:SUSY} lead to the stationary points of the scalar potential in four-dimensional ${\cal N}=1$ supergravity in general. 
In this respect, we focus on the fixed point of $\tau$ which will be a candidate of the solution \eqref{eq:vev_sp4z1}. Remarkably, among all the fixed points of $\tau$ in Table \ref{tab:Sp6Zfp}, we find that only the following fixed point leads to $Y_{\mathbf{1}}^{(4)}(\tau_{\rm fix})=0$:
\begin{align}
\tau_{\rm fix} =\frac{i\omega^2}{\sqrt{3}}
\left(
\begin{array}{ccc}
  -\frac{1+i 3\sqrt{3}}{4} & \frac{1}{2}\omega &  \omega^2 \\
  \frac{1}{2}\omega & -\frac{1+i3\sqrt{3}}{4} &  \omega^2 \\
  \omega^2 & \omega^2 &  \omega \\
\end{array}
\right).
\label{eq:fix_Sp6Z}
\end{align}
Since \eqref{eq:fix_Sp6Z} is a single constraint on the moduli, 
only one of the moduli becomes massive. 
By calculating the mass matrix of the moduli, we find that the following 
linear combination is indeed stabilized:
\begin{align}
    -\frac{1}{2}(\tau_1 + \tau_2) + \tau_3 - \tau_4 + \tau_5 + \tau_6,
\end{align}
and other directions are massless. 
Here, we use
\begin{align}
    &\langle \partial_{\tau_1} Y_{\mathbf{1}}^{(4)}(\tau)\rangle = \langle \partial_{\tau_2} Y_{\mathbf{1}}^{(4)}(\tau)\rangle \simeq 3.18i,
    \nonumber\\
    &\langle \partial_{\tau_3} Y_{\mathbf{1}}^{(4)}(\tau)\rangle = - \langle \partial_{\tau_4} Y_{\mathbf{1}}^{(4)}(\tau)\rangle = \langle \partial_{\tau_5} Y_{\mathbf{1}}^{(4)}(\tau)\rangle = \langle \partial_{\tau_6} Y_{\mathbf{1}}^{(4)}(\tau)\rangle \simeq - 6.36i.
\end{align}
We also examine whether other singlet modular forms with modular weights up to five are zero at all the fixed points. 
It can be verified by the modular transformation of the singlet modular form $Y_{\mathbbm1}^{(k_Y)}$:
\begin{align}
    Y_{\mathbf{1}}^{(k_Y)}(h_\ast \tau) = \bigl[{\rm det}(C \tau +D)]^{k_Y} Y_{\mathbf{1}}^{(k_Y)}(\tau),
\end{align}
where
\begin{align}
    h_\ast =
    \begin{pmatrix}
        A & B\\
        C & D
    \end{pmatrix}
    =
    \begin{pmatrix}
       0  &  \mathbbm1_3\\
      -\mathbbm1_3   & {\cal C}_1
    \end{pmatrix}
\end{align}
with
\begin{align}
   {\cal C}_1 =
\begin{pmatrix}
  1 & 0 & 1   \\
  0 & 1 & 1  \\
  1 & 1 & -1  \\
\end{pmatrix}
\end{align}
being the generator of the stabilizer group at the fixed point \eqref{eq:fix_Sp6Z} in Table \ref{tab:Sp6Zfp}. 
It is then found that \eqref{eq:Ymodulartrf} obeys $ \bigl[{\rm det}(C \tau +D)]^{k_Y}\neq 1$ for $k_Y=2,7/2, 4, 5$. 
Hence, the modular forms of weight 2, 7/2, 4, 5 are zero at the fixed point \eqref{eq:fix_Sp6Z}.\footnote{The singlet modular forms up to the weight five are listed in Appendix \ref{app:sp6modular}.}
Note that there are two modular forms of weight 4 both of which is zero at the fixed point \eqref{eq:fix_Sp6Z}. 
Hence, our scenario is also applicable to these modular forms.

In the following, 
we discuss the possibility of stabilizing the other moduli fields by using the matter superpotential. 
Before going to the detail about the moduli stabilization, let us analyze the structure of singlet modular forms under $\tilde{\Delta}(96)$. 
As shown in Appendix \ref{app:sp6modular}, all the singlet modular forms are constructed from three fundamental modular functions $\{ \zeta(\tau), \eta(\tau), \theta(\tau)\}$ in \eqref{eq:zeromodes}. 
Specifically, three fundamental modular functions are rewritten as
\begin{align}
    \zeta(\tau) &= \sqrt{2}\sum_{\vec{m}\in\mathbb{Z}^3}
    e^{\pi i(\vec{m}+a_1)^{t}\tau^{\prime}(\vec{m}+a_1)},
    \nonumber\\
    \eta(\tau) &= \sum_{\vec{m}\in\mathbb{Z}^3}
    e^{\pi i(\vec{m}+a_2)^{t}\tau^{\prime}(\vec{m}+a_2)},
    \nonumber\\
    \theta(\tau)&=\sum_{\vec{m}\in\mathbb{Z}^3}
    e^{\pi i(\vec{m}+a_3)^{t}\tau^{\prime}(\vec{m}+a_3)},
\end{align}
with
\begin{align}
    a_1=\frac{1}{4}\begin{psmallmatrix}
        1\\ 3\\ 2\\
    \end{psmallmatrix},\qquad
    a_2=\begin{psmallmatrix}
        0\\0\\0\\
    \end{psmallmatrix},\qquad
    a_3=\frac{1}{2}\begin{psmallmatrix}
        1\\1\\0\\
    \end{psmallmatrix}
    .
\end{align}
Remarkably, they have the following properties. 
$\eta(\tau)$ is invariant under the following three symmetries: i) an exchange of $\{\tau_1, \tau_4, m_1\} \rightarrow \{\tau_3, \tau_6, m_3\}$, ii) an exchange of $\{\tau_2, \tau_4, m_2\} \rightarrow \{\tau_3, \tau_5, m_3\}$, and iii) an exchange of $\{\tau_1, \tau_5, m_1\} \rightarrow \{\tau_2, \tau_6, m_2\}$. Here, we denote $\vec{m} = (m_1, m_2, m_3)^t$. In addition, $\zeta(\tau)$ and $\theta(\tau)$ are invariant under an exchange of $\{\tau_1, \tau_5, m_1\} \rightarrow \{\tau_2, \tau_6, m_2\}$. 
Since the singlet modular form is a non-trivial product of these fundamental modular functions $\{ \zeta(\tau), \eta(\tau), \theta(\tau)\}$, we conclude that the singlet modular forms are only invariant under the exchange of $\{\tau_1, \tau_5\} \rightarrow \{\tau_2, \tau_6\}$.

To stabilize the multiple moduli fields, we extend the previous simple superpotential \eqref{eq:KWsp6sim} to 
\begin{align}
    K &= -h \ln \left( \det (-i\tau + i \Bar{\tau})\right) + \sum_{i=1}^5 \frac{|X_i|^2}{\det (-i\tau + i \Bar{\tau})^{k_i}}\,,
    \nonumber\\
    W &= Y_{\mathbf{1}}^{(2)} X_1 + Y_{\mathbf{1}}^{(7/2)} X_2 +  Y_{\mathbf{1}}^{(4_a)} X_3 +  Y_{\mathbf{1}}^{(4_b)} X_4 + Y_{\mathbf{1}}^{(5)} X_5,
\label{eq:KWsp6}
\end{align}
where we introduce five matter superfields $X_i$ with the modular weight $k_i$:
\begin{align}
    k_1 = 2,\quad
    k_2 = 7/2,\quad    
    k_3 = k_4 = 4,\quad
    k_5 = 5.
\end{align}
Note that all the Siegel modular forms of $Y_{\mathbf{1}}^{(k_Y)}$ shown above are zero at the fixed point \eqref{eq:fix_Sp6Z}. 
Hence, the matter $X$ and moduli $\tau_{n}$ with $n=1,2,...,6$ can be stabilized at
\begin{align}
    \langle Y_{\mathbf{1}}^{(k_{Y_{i}})} (\tau)\rangle = \langle X \rangle = 0,
\end{align}
with $k_Y=\{2, 7/2, 4_a, 4_b, 5\}$. 
As in the $Sp(4,\mathbb{Z})$ case, let us count the number of massive moduli fields 
by expanding the superpotential \eqref{eq:KWsp6} up to the second order of $\delta X= X- \langle X\rangle$ and $\delta \tau_{n} = \tau_{n} - \langle\tau_{n}\rangle$ with $n=1,2,...,6$:
\small
\begin{align}
    W = \sum_{i=1}^5 \delta X_i \left( \frac{\partial Y_{\mathbf{1}}^{(k_{Y_i})}}{\partial \tau_1} \delta \tau_1 +
    \frac{\partial Y_{\mathbf{1}}^{(k_{Y_i})}}{\partial \tau_2} \delta \tau_2 +
    \frac{\partial Y_{\mathbf{1}}^{(k_{Y_i})}}{\partial \tau_3} \delta \tau_3 +
    \frac{\partial Y_{\mathbf{1}}^{(k_{Y_i})}}{\partial \tau_4} \delta \tau_4+
    \frac{\partial Y_{\mathbf{1}}^{(k_{Y_i})}}{\partial \tau_5} \delta \tau_5+
    \frac{\partial Y_{\mathbf{1}}^{(k_{Y_i})}}{\partial \tau_6} \delta \tau_6\right).
\end{align}
\normalsize
Remarkably, the first derivatives of some modular forms with respect to moduli fields 
at the fixed point \eqref{eq:fix_Sp6Z} are related to each other:
\begin{align}
    \frac{\partial}{\partial \tau_n}Y^{\left(\frac{7}{2}\right)}_{\mathbf{1}}\biggl|_{\tau=\tau_{\rm fix}}&=
    Y^{\left(\frac{3}{2}\right)}_{\mathbf{1}}\frac{\partial}{\partial \tau_n}Y^{(2)}_{\mathbf{1}}\biggl|_{\tau=\tau_{\rm fix}},
    \nonumber\\
    \frac{\partial}{\partial \tau_n}Y^{(4_b)}_{\mathbf{1}}\biggl|_{\tau=\tau_{\rm fix}} &=
    2Y^{(2)}_{\mathbf{1}}\frac{\partial}{\partial \tau_n}Y^{(2)}_{\mathbf{1}}\biggl|_{\tau=\tau_{\rm fix}}=0,
    \nonumber\\
    \frac{\partial}{\partial \tau_n}Y^{(5)}_{\mathbf{1}}\biggl|_{\tau=\tau_{\rm fix}}&=(\zeta^2\theta^4+\zeta^2\eta^4-\zeta^2\theta^2\eta^2)\frac{\partial}{\partial \tau_n}Y^{(2)}_{\mathbf{1}}\biggl|_{\tau=\tau_{\rm fix}},
\end{align}
with $n=1,2,...,6$, where we use 
\begin{align}
    Y^{(2)}_{\mathbf{1}}\bigl|_{\tau=\tau_{\rm fix}}&=\frac{2\sqrt{3}}{3}\eta\theta^3+\frac{2\sqrt{3}}{3}\eta^3\theta+\frac{1}{\sqrt{3}}\zeta^4\bigl|_{\tau=\tau_{\rm fix}}=0,
    \nonumber\\
    Y^{\left(\frac{7}{2}\right)}_{\mathbf{1}}\bigl|_{\tau=\tau_{\rm fix}}&=Y^{(2)}_{\mathbf{1}}Y^{\left(\frac{3}{2}\right)}_{\mathbf{1}}\bigl|_{\tau=\tau_{\rm fix}}=0,
    \nonumber\\
    Y^{(4_b)}_{\mathbf{1}}\bigl|_{\tau=\tau_{\rm fix}}&=Y^{(2)}_{\mathbf{1}}Y^{(2)}_{\mathbf{1}}\bigl|_{\tau=\tau_{\rm fix}} =0,
    \nonumber\\
    Y^{(5)}_{\mathbf{1}}\bigl|_{\tau=\tau_{\rm fix}}&=Y^{(2)}_{\mathbf{1}}(\zeta^2\theta^4+\zeta^2\eta^4-\zeta^2\theta^2\eta^2)\bigl|_{\tau=\tau_{\rm fix}}=0.
\end{align}
Hence, $Y_{\mathbf{1}}^{(k)}$ with $k=7/2, 4_b, 5$ is proportional to $Y^{(2)}_{\mathbf{1}}$ and their first derivatives with respect to moduli fields are proportional to $\frac{\partial}{\partial \tau_n}Y^{(2)}_{\mathbf{1}}$ at the fixed point \eqref{eq:fix_Sp6Z}. 
It turns out that there are only two directions to be fixed using these modular forms. 
The stabilization of all the moduli fields could be achieved by introducing Coleman-Weinberg potential with multiple vector-like pairs as well as non-perturbative effects for axion-like fields, but beyond the scope of this paper.

\subsection{Siegel modular forms of $Sp(6,\mathbb{Z})$}
\label{app:sp6modular}

In this appendix, we summarize the singlet modular forms of $\tilde{\Delta}(96)$ in the convention of Ref. \cite{Kikuchi:2023dow}. 
By using the Riemann-theta function in \eqref{eq:zeromodes}, 
the singlet modular forms up to the weight five are given by

\begin{align}
     Y^{(\frac{3}{2})}_{\mathbf{1}} &=\sqrt{\frac{3}{2}}\zeta(\eta-\theta)(\eta+\theta),
     \nonumber\\
     Y^{(2)}_{\mathbf{1}} &=\frac{1}{\sqrt{3}}(\zeta^4+2\eta^3\theta+2\eta\theta^3),
     \nonumber\\
     Y^{(3)}_{\mathbf{1}}&=\frac{3}{2}\zeta^2(\eta-\theta)^2(\eta+\theta)^2,
     \nonumber\\
     Y^{(\frac{7}{2})}_{\mathbf{1}}&=\frac{1}{\sqrt{2}}\zeta(\eta-\theta)(\eta+\theta)(\zeta^4+2\eta^3\theta+2\eta\theta^3),
     \nonumber\\
       Y^{(4_a)}_{\mathbf{1}}&=\frac{1}{4\sqrt{6}}(\eta^8-32\zeta^4\eta^3\theta-4\eta^6\theta^2-32\zeta^4\eta\theta^3+6\eta^4\theta^4-4\eta^2\theta^6+\theta^8),
       \nonumber\\
        Y^{(4_b)}_{\mathbf{1}}&=\frac{1}{3}(\zeta^4+2\eta^3\theta+2\eta\theta^3)^2
        \nonumber\\
        Y^{(\frac{9}{2})}_{\mathbf{1}}&=-\frac{1}{2}\sqrt{\frac{3}{2}}\zeta^3(\eta-\theta)^3(\eta+\theta)^3,
        \nonumber\\
         Y^{(5)}_{\mathbf{1}}&=\frac{1}{2\sqrt{3}}\zeta^2(\eta-\theta)^2(\eta+\theta)^2(\zeta^4+2\eta^3\theta+2\eta\theta^3).
\end{align}
 We find that the vanishing modular forms up to the weight 5 can be achieved by three conditions: $\zeta^4+2\eta^3\theta + 2\eta \theta^3=0$, $Y^{(4_b)}_{\mathbf{1}}=0$ and $\eta=\theta$ or $\eta=-\theta$. Hence, the maximum number of moduli fields that are stabilized at the points with vanishing modular forms is three when we utilize the modular forms up to the weight 5.

\bibliography{references}{}
\bibliographystyle{JHEP}

\end{document}